\documentclass[pdflatex,sn-mathphys]{sn-jnl}

\jyear{2021}
\theoremstyle{thmstyleone}

\theoremstyle{thmstyletwo}%

\theoremstyle{thmstylethree}%
\usepackage{amsbsy}
\usepackage{wrapfig}
\usepackage{framed}
\normalbaroutside

\begin{document}

\title[Better than DFA]{Better than DFA? A Bayesian Method for Estimating the Hurst Exponent in Behavioral Sciences}

\author*[1]{\fnm{Aaron D.} \sur{Likens}}\email{alikens@unomaha.edu}

\author[1]{\fnm{Madhur} \sur{Mangalam}}\email{mmangalam@unomaha.edu}

\author[2]{\fnm{Aaron Y.} \sur{Wong}}\email{wonga@umn.edu}

\author[1]{\fnm{Anaelle C.} \sur{Charles}}\email{anaellecharles@unomaha.edu}

\author[2]{\fnm{Caitlin} \sur{Mills}}\email{cmills@umn.edu}

\affil[1]{\orgdiv{Division of Biomechanics and Research Development, Department of Biomechanics, and Center for Research in Human Movement Variability}, \orgname{University of Nebraska at Omaha}, \orgaddress{\street{6160 University Dr S}, \city{Omaha}, \postcode{68182}, \state{NE}, \country{USA}}}

\affil[2]{\orgdiv{Department of Educational Psychology}, \orgname{University of Minnesota},\\ \orgaddress{\street{56 East River Road}, \city{Minneapolis}, \postcode{55415}, \state{MN}, \country{USA}}}

\abstract{Detrended Fluctuation Analysis (DFA) is the most popular fractal analytical technique used to evaluate the strength of long-range correlations in empirical time series in terms of the Hurst exponent, $H$. Specifically, DFA quantifies the linear regression slope in log-log coordinates representing the relationship between the time series' variability and the number of timescales over which this variability is computed. We compared the performance of two methods of fractal analysis---the current gold standard, DFA, and a Bayesian method that is not currently well-known in behavioral sciences: the Hurst-Kolmogorov (HK) method---in estimating the Hurst exponent of synthetic and empirical time series. Simulations demonstrate that the HK method consistently outperforms DFA in three important ways. The HK method: (i) accurately assesses long-range correlations when the measurement time series is short, (ii) shows minimal dispersion about the central tendency, and (iii) yields a point estimate that does not depend on the length of the measurement time series or its underlying Hurst exponent. Comparing the two methods using empirical time series from multiple settings further supports these findings. We conclude that applying DFA to synthetic time series and empirical time series during brief trials is unreliable and encourage the systematic application of the HK method to assess the Hurst exponent of empirical time series in behavioral sciences.}


\maketitle

\section{Introduction}

Behavior in humans is fluid. Repetitions of gross movements, such as walking, and fine movements, such as tapping a finger, vary from one cycle to the next. Even the most basic behavioral measurement---the reaction time---ebbs and flows around a typical value, the arithmetic $Mean$. The standard deviation, $SD$, measures the average distance of each point from that $Mean$ and carries the assumption that deviations from the $Mean$ are errors surrounding an intended stride or a tapping response. Decades of research refute this assumption in the serial measurement of human behavior. The ``variability is error" assumption means that behavioral measurements should be independent of one observation to the next. However, an inspection of temporal sequences of measurements reveals that behaviors correlate with one another over time. Long steps tend to follow long steps; fast responses tend to follow fast responses. The closer in time, the closer the resemblance. Conversely, correlation decays with greater separation in time. The quantification of these long-range relationships is, therefore, of critical importance in behavioral sciences. However, long-range correlations are not amenable to measurement by descriptive statistics such as $SD$, coefficient of variation ($CoV$), and root mean square ($RMS$).

\begin{figure*}
\begin{center}
\includegraphics[width=0.75\textwidth]{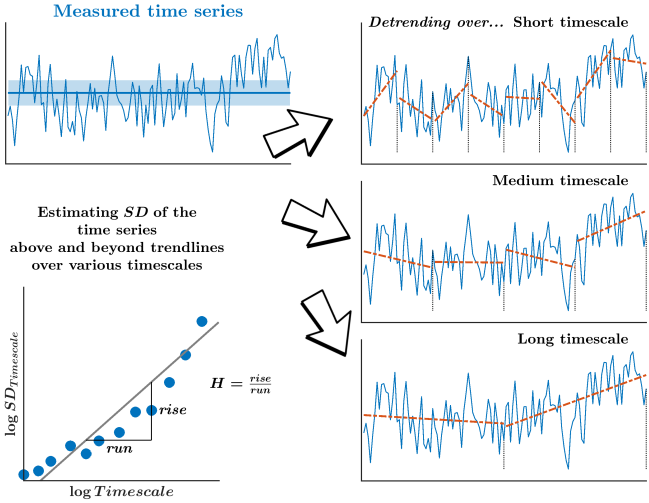}
\caption{\textbf{Schematic portrayal of the measure of fractality, \boldmath$H$, yielded by the DFA.} $H$ relates how the $SD$-like variation grows across many timescales, statistically encoding how the correlation among sequential measurements might decay slowly across longer separations in time. We use detrending of these variations over progressively longer timescales to remove the mean drift across each of these timescales.}
\label{fig:HfGn}
\end{center}
\end{figure*}

A robust approach to assessing how long-range correlations between measurements decline over longer time intervals is to use the Hurst exponent, $H$ \cite{hurst1951long}. The Hurst exponent, $H$, was named by Mandelbrot \cite{mandelbrot1969computer} in honor of pioneering work by Edwin Hurst in the field of hydrology, the ``fractal" flood characteristics of the Nile River delta \cite{hurst1951long}. According to Mandelbrot, $H$ measures the presence of long-run statistical persistence in a time series, as well as its intensity \cite{beran1994estimation,mandelbrot1969computer}. $H$ describes how the measurements' $SD$-like variations grow across progressively longer timescales, indicating the rate at which correlation among sequential measurements decay across subsequent separations in time (\textbf{Fig. \ref{fig:HfGn}}). More precisely, the Hurst exponent describes a single fractal-scaling estimate of power-law decay in the autocorrelation $\rho$ for lag $k$ as $\rho_{k} = |k + 1|^{2H} -2|k|^{2H} + |k - 1|^{2H}$, for which $H$ reveals the presence and degree of persistent correlations ($0.5 < H < 1.0$, wherein large values are typically followed by large values and vice versa) or anti-persistent correlations ($0 < H < 0.5$), wherein small values typically follow large values and vice versa. An empirical time series with $H \rightarrow 0.5$ implies a random process where subsequent observations are uncorrelated.  

Detrended fluctuation analysis (DFA) is the most used technique to uncover long-range correlations in diverse research fields, such as material science \cite{kantelhardt1999phases}, meteorology \cite{efstathiou2010altitude,ivanova1999application,tatli2020long}, economics \cite{alvarez2008short,grau2000empirical,ivanov2004common,liu1997correlations,liu1999statistical}, ethology \cite{alados2000fractal,bee2001individual}, bioinformatics \cite{buldyrev1998analysis,mantegna1994linguistic,peng1993finite}, and physiology \cite{castiglioni2019fast,goldberger2002fractal,hardstone2012detrended,peng1993long}. The Hurst exponent estimated using DFA has also proved to be extremely powerful in its capacity to uncover system dynamics, such as feedforward and forward processes in postural control \cite{delignieres2011transition,duarte2008complexity,lin2008reliability}, system-wide coordination in motor control \cite{chen1997long,diniz2011contemporary}, cognition \cite{allegrini2009spontaneous,gilden1995noise,kello2010scaling,stephen2008strong,van2003self}, and perception-action \cite{mangalam2019fractal,mangalam2020bodywide,mangalam2020global}. The Hurst exponent estimated using DFA also helps identify different states of the same system according to its different scaling behaviors, for instance, the $H$ values for heart interbeat intervals between healthy individuals and those with disease \cite{ashkenazy1999discrimination,ho1997predicting,peng1995fractal}. Likewise, the $H$ values for stride intervals during walking are different for healthy adults and individuals with movement deficits due to aging and pathology \cite{bartsch2007fluctuation,hausdorff1997altered,hausdorff2001human,hausdorff2007gait,herman2005gait,kobsar2014evaluation}. The Hurst exponent, typically estimated using DFA, also serves as a critical benchmark for developing interventions \cite{mangalam2022leveraging,raffalt2021temporal,raffalt2023stride} and quantifying their effects \cite{kaipust2013gait,marmelat2020fractal,vaz2020gait}. In short, DFA has become central to quantifying the Hurst exponent across diverse research fields, including behavioral sciences.

The most significant advantage of DFA over other methods of assessing the strength of long-range correlations in empirical time series is that it is suitable for nonstationary time series, thereby preventing erroneous detection of long-range correlations that are a side effect of non-stationarity. However, the Hurst exponent yielded by the DFA becomes unstable because of the nonlinear filtering characteristics associated with detrending \cite{kiyono2016nonlinear}. Therefore, DFA has been modified by introducing different detrending techniques, such as the centered moving average (CMA) method \cite{alvarez2005detrending}, detrended moving average (DMA) method \cite{arianos2007detrending}, the modified detrended fluctuation analysis (MDFA) \cite{holl2019theoretical}, and orthogonal detrended fluctuation analysis \cite{govindan2020detrended}. Different detrending methods show various advantages and limitations, depending on the presence of long-range trends \cite{chen2005effect,hu2001effect}. For instance, CMA is slightly superior to the original DFA algorithm in terms of straighter fluctuation curves \cite{shao2012comparing}, and DFA based on empirical mode decomposition (EMD) is superior to the traditional DFA when the time series is strongly anticorrelated \cite{qian2011modified}. DMA method is superior to the traditional DFA for time series with $0.2 < H < 0.8$, while traditional DFA performs better when $H > 0.8$ \cite{xu2005quantifying}. Numerical analysis shows the traditional DFA still confers several advantages, mainly when the data trend's functional form is not known \textit{a priori} \cite{bashan2008comparison,grech2005statistical}.

Nonetheless, DFA has several shortcomings beyond the detrending procedure. For instance, numerous authors have pointed out that DFA does not accurately assess long-range correlations when the empirical time series is short \cite{dlask2019hurst,katsev2003hurst,schaefer2014comparative}, producing a positive bias in its central tendency in addition to a large dispersion \cite{almurad2016evenly,delignieres2006fractal,marmelat2019fractal,ravi2020assessing,roume2019biases,yuan2018unbiased}. Often an empirical time series with more than $500$ samples is required to use DFA with reasonable accuracy. This requirement is a significant limitation, especially when it is impractical to collect a long measurement time series due to time constraints and financial or clinical reasons \cite{marmelat2019fractal}. In addition, many cognitive and psychological phenomena are fleeting and ephemeral such as moments of insight \cite{stephen2009dynamics,stephen2012scaling}. As it stands, the outcome yielded by many other methods of assessing the strength of long-range correlations in measurement time series is precariously sensitive to the length of the measurement time series. However, DFA generally performs best \cite{delignieres2006fractal,stroe2009estimating}. Therefore, there is an urgent need in behavioral sciences for an analytical method that: (i) accurately assesses long-range correlations when the measurement time series is short, (ii) shows minimal dispersion about the central tendency, and (iii) yields a point estimate that does not depend on the length of the measurement time series or its underlying Hurst exponent. No such methods are currently widely used, thus limiting our ability to make strong inferences in those many limiting domains noted above.

In this paper, we present a simulation study comparing two methods of fractal analysis, the current gold standard, DFA \cite{peng1994mosaic,peng1995quantification}, and a Bayesian method that is not well-known in behavioral sciences---the Hurst-Kolmogorov (HK) methodology \cite{tyralis2014bayesian}. We use these simulation results to inform four empirical human behavioral time series analyses. Those studies capture a broad swath of common behavioral measurements---gait, sensorimotor synchronization, and reaction times---derived from tasks typically conceived as purely motor and those considered more purely cognitive. Using synthetic and empirical time series, we show that the HK method outperforms DFA in all three benchmarks described above.

\section{Two methods of estimating the Hurst exponent}

\subsection{Estimating the Hurst exponent using the HK method}

Tyralis and Koutsoyiannis \cite{tyralis2014bayesian} developed a Bayesian method for estimating $H$. As we will show, this method offers a viable solution to several issues with DFA outlined above. As a preview, we show that the HK method outperforms DFA across a broad range of $H$, especially when time series are short. In the remainder of this section, we overview the HK method. Additional details, including mathematical proofs, can be found in Tyralis and Koutsoyiannis \cite{tyralis2014bayesian}. In this description, we generally follow their notation.

Koutsoyiannis \cite{koutsoyiannis2003climate} report that the autocorrelation function for the so-called Hurst-Kolmogorov (HK) process is given by:

\begin{align*}
\rho_{k} = |k + 1|^{2H}/2 -2|k|^{2H}/2 + |k - 1|^{2H}, \quad k = 0,1,\dots, \tag{1}\label{eq:1}
\end{align*}

\noindent{}where $H$ is the Hurst exponent, $k$ is the time lag, and $\rho_{k}$ is the autocorrelation for a given $k$. When $H = 0.5$, $\rho_{k}$ is zero for all $k > 0$ but 1 when $k=0$. When $0 < H < 0.5$, $\rho_{k}$ is negative at lag $1$ but damps towards zero for $k > 1$; when $0.5 < H < 1$, $\rho_{k}$ is positive at lag 1 but slowly decays to zero; and as $H \rightarrow 1$, $\rho_{k}$ approaches $0$ asymptotically. 

Tyralis and Koutsoyiannis \cite{tyralis2014bayesian} employ a Bayesian technique for estimating the Hurst exponent. In that work, they derive a method to sample from the posterior distribution of $H$ that takes the following form:

\begin{align*}
\pi(\boldsymbol{\varphi}|\textbf{x}_{n}) \propto |\textbf{R}_{n}|^{-1/2} \: [\textbf{e}_{n}^{T} \textbf{R}_{n}^{-1} \textbf{e}_{n} \textbf{x}_{n}^{T} \textbf{R}_{n}^{-1} \textbf{x}_{n} - (\textbf{e}_{n}^{T} \textbf{R}_{n}^{-1} \textbf{e}_{n})^{2}]^{-(n-1)/2}\\ (\textbf{e}_{n}^{T} \textbf{R}_{n}^{-1} \textbf{e}_{n})^{n/2 - 1}, \tag{2}\label{eq:2}
\end{align*}

\noindent{}and its natural logarithm is then:

\begin{align*}
\ln{\pi(\boldsymbol{\varphi}|\textbf{x}_{n})} \propto \frac{1}{2} \ln{|\textbf{R}_{n}|} \: -\frac{(n-1)}{2} \ln{[\textbf{e}_{n}^{T} \textbf{R}_{n}^{-1} \textbf{e}_{n} x_{n}^{T} \textbf{R}_{n}^{-1} \textbf{x}_{n} - (\textbf{e}_{n}^{T} \textbf{R}_{n}^{-1} \textbf{e}_{n})^{2}]}\\ +
\frac{n - 2}{2} \ln{(\textbf{e}_{n}^{T} \textbf{R}_{n}^{-1} \textbf{e}_{n})}, \tag{3}\label{eq:3}
\end{align*}

\noindent{}where $\textbf{R}_{n}$ is the autocorrelation matrix with elements $r_{i,j}$ where $i, j = 1, 2, 3, \dots, n$, $\textbf{e}_{n} = (1, 1, 1, \dots, 1)^{T}$ is a vector of ones with $n$ elements, $|\dots|$ indicates a determinant, the superscript in $\textbf{R}_{n}^{-1}$ indicates a matrix inverse, and the superscript $T$ indicates a matrix transpose. The matrix products on the right-hand side of Eq.~\ref{eq:3} are built from the quadratic forms for the inverse of a symmetric, positive definite autocorrelation matrix which can be obtained using the Levinson algorithm (Algorithm 4.7.2, Golub \& Van Loan \cite{golub2013matrix}, p. 235) for a given $x_{t} \rho_{k}$. 

Accept-reject algorithms are standard, powerful tools for sampling from complex distributions and follow a simple set of steps \cite{robert1999monte}. Suppose a probability density function (PDF) exists, $f(x)$, from which it is difficult to sample. We refer to $f(x)$ as the target distribution. One can use the Monte Carlo method to sample from $f(x)$. The algorithm is as follows. First, one samples from a simpler proposal distribution from which it is easy to sample, $Mg(x)$, where $g(x)$ has the same domain as $f(x)$ and $M$ is a constant large enough such that $g(x) \geq f(x)$. The proposal PDFs can take many forms, such as uniform or truncated Gaussian distributions. Computational efficiency is gained if the overall shape of $g(x)$ is similar to $f(x)$. Second, one evaluates $f(x)$ at the value proposed by sampling from $g(x)$. Third, one draws a sample from the $U(x) \sim Uniform(0, Mg(x))$. If $U(x) \leq f(x)$, then we accept the proposed value from $g(x)$ as a valid sample. Otherwise, we reject the proposal from $g(x)$. This process is repeated for $n$ samples, where $n$ is the number of samples we wish to draw from the posterior distribution.

In the present case, we used the accept-reject algorithm to sample from the posterior distribution of $H$ (Algorithm A.5, Robert \& Casella \cite{robert1999monte}, p. 49). The target distribution, $f(x)$ is Eq.~\ref{eq:3} and $g(x) \sim Uniform(0,1)$. The choice of $g(x)$ makes sense in this case because $g(x)$ shares the same domain of $H$ and hence Eq.~\ref{eq:3}, namely $(0,1)$ \cite{tyralis2014bayesian}. $M$ is chosen using a numerical optimization routine that finds the maximum of Eq.~\ref{eq:3} as a function of $H$. Finally, from the sampled posterior distribution of $H$, we take the median of the distribution as a point estimate of $H$. Time series were submitted to the HK method using $R$ \cite{team2013r} using the function \texttt{inferH()} from the package ``HKprocess" \cite{tyralis2022hkpackage}. The function \texttt{inferH()} has two inputs: the time series, $x_{N}$, and the size of the simulated sample from the posterior distribution of $H$, $n$. We set the $n$ to $500$.

\subsection{Estimating the Hurst exponent using DFA}

We used DFA---as described by Peng et al. \cite{peng1994mosaic,peng1995quantification}---to access the strength of long-range correlations in synthetic time series of different \textit{a priori} known values of $H$ and empirical human behavioral time series. DFA computes the Hurst exponent, $H$, using the first-order integration of time series $x_{t}$ of length $N$, where $t \in \mathbb{N}$:

\begin{align*}
X_{t} = \sum_{i=1}^{N} (x_{i} - \langle x \rangle), \tag{4}\label{eq:4}
\end{align*}

\noindent{}where $\langle x \rangle$ is the grand mean of the time series. It computes root mean square ($RMS$; that is, averaging the residuals) for each linear trend $Y_{t}$ fit to non-overlapping $n$-length bins to build fluctuation function:

\begin{align*}
f(n) = \sqrt{\frac{1}{N} \sum_{t=1}^{N} (X_{t} - Y_{t})}, \tag{5}\label{eq:5}
\end{align*}

\noindent{}for $n<N/4$. $f(n)$ is a power law:
\begin{align*}
f(n) \sim n^{H}, \tag{6}\label{eq:6}
\end{align*}

\noindent{}where $H$ is the Hurst exponent estimable using logarithmic transformation:

\begin{align*}
H = \frac{\ln{f(n)}}{\ln{n}}. \tag{7}\label{eq:7}
\end{align*}

A bin size range of $[4, N/2]$ was used for the DFA in the present study, which is standard practice while using DFA \citep{damouras2010empirical,farag2013automated,jordan2006long,likens2015experimental,likens2020tutorial}. Time series were submitted to the DFA in \texttt{R} \citep{team2013r} using the function \texttt{dfa()} from the package ``fractalRegression" \cite{likens2021fractalregression}.

The computational details of the two methods are relatively distinct, with the HK method having its foundations in the Bayes theorem, whereas DFA computes the Hurst exponent directly from the time series data.

\section{Simulations}

\subsection{Methods}

We used the Davies-Harte algorithm \cite{davies1987tests} to generate synthetic time series of varying lengths ($N = 32, 64, 128, 256, 512, 1024$) and varying values of the Hurst exponent ($H = 0.1, 0.2, \dots, 0.9$). This algorithm generates fractional Gaussian noise (fGn), which has been proposed as a model to understand the long-range correlations postulated to occur in various behavioral systems \cite{allegrini2009spontaneous,chen1997long,diniz2011contemporary,gilden1995noise,gilden2001cognitive,grigolini2009theory,kello2010scaling,van2003self,van2005human}. We generated $1,000$ synthetic time series for each combination of $N$ and $H$ in $R$ \citep{team2013r} using the function \texttt{fgn\_sim()} from the package ``fractalRegression" \cite{likens2021fractalregression} and submitted them to the HK method and DFA.

\subsection{Results}

\begin{figure*}
\centering
\includegraphics[width=0.75\textwidth]{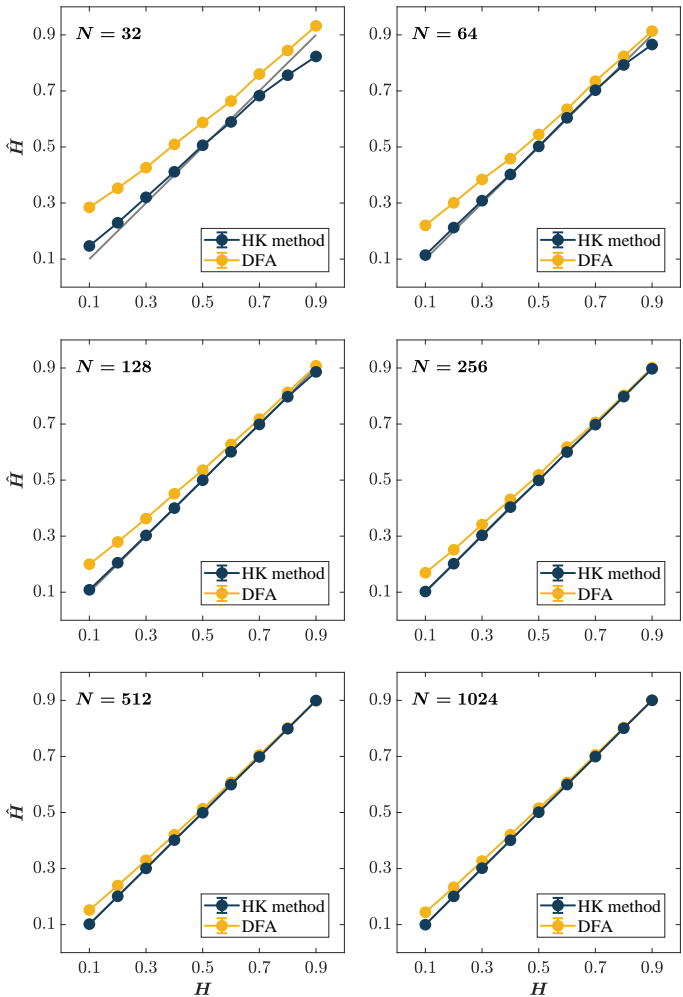}
\caption{\textbf{The HK method estimates the Hurst exponent, \boldmath$\hat{H}$, with consistently better accuracy than DFA, which overestimates \boldmath$\hat{H}$, specifically for short time series and small values of \boldmath$H$.} Each panel plots the $Mean$ estimated values of $\hat{H}$ for $1,000$ synthetic time series of length $N = 32, 64, 128, 256, 512, 1024$ with \textit{a priori} known values of $H$. The grey line indicates the ideal case where the estimated value is the same as the actual value, i.e., $\hat{H} = H$. Error bars indicate $95$\% CI across $1000$ simulations.}
\label{fig:DFAvsHKp_1}
\end{figure*}

\begin{figure*}
\begin{center}
\includegraphics[width=0.75\textwidth]{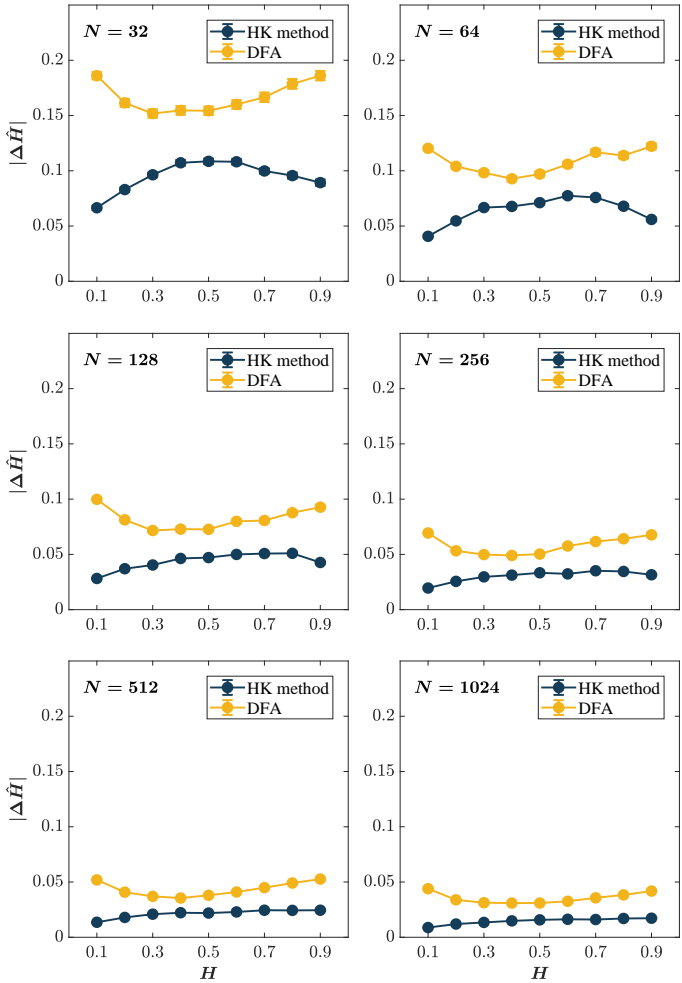}
\caption{\textbf{Although DFA estimates the Hurst exponent, \boldmath$\hat{H}$, reasonably accurately for long time series (\boldmath$|\Delta \hat{H}| \sim 0.05$ for \boldmath$N > 512$), the HK method estimates \boldmath$H$ with consistently better accuracy than DFA.} Each panel plots the $Mean$ absolute error in the estimation of $\hat{H}$, $|\Delta \hat{H}|$, for $1,000$ synthetic time series of length $N = 32, 64, 128, 256, 512, 1024$ with \textit{a priori} known values of $H$.  Error bars indicate $95$\% CI across $1000$ simulations.}
\label{fig:DFAvsHKp_2}
\end{center}
\end{figure*}

\textbf{Figs. \ref{fig:DFAvsHKp_1} \& \ref{fig:DFAvsHKp_2}} provide a summary visualization of the simulation results for each combination of the time series lengths ($N = 32, 64, \dots, 1024$), and the \textit{a priori} known values of the Hurst exponent ($H = 0.1, 0.2, \dots, 0.9$). As a general preview, in all one but the shortest time series, $N = 32$, where neither method was useful, the HK method outperforms DFA in estimating $\hat{H}$ (\textbf{Fig. \ref{fig:DFAvsHKp_1}}). For the shortest time series, $N = 32$, both methods produce unreasonable errors ($Mean$ $|\Delta \hat{H}| > 0.05$; \textbf{Fig. \ref{fig:DFAvsHKp_2}, top left}), although the HK method is still somewhat unbiased in its central tendency, producing $Mean$ $\hat{H}$ close to the \textit{a priori} known values of $H$ (\textbf{Fig. \ref{fig:DFAvsHKp_1}, top right}).

When $N = 64$, a very short time series compared to the DFA standard of $> 500$, the HK method and DFA show considerable differences in performance. First, the $Mean$ $\hat{H}$ estimated by the HK method closely approximates the \textit{a priori} known values of $H$, while DFA produces substantially and uniformly positive bias in mean $\hat{H}$ across the entire range of $H$ (\textbf{Fig. \ref{fig:DFAvsHKp_1}, top right}). Second, the HK method produces substantially smaller $Mean$ absolute errors, specifically $|\Delta \hat{H}|$ falling within a range that could be used, with caution, in analyzing short time series (\textbf{Fig. \ref{fig:DFAvsHKp_2}, top left}). When $N = 128$, $Mean$ $\hat{H}$ are virtually indistinguishable from nominal values, while DFA remains positively biased (\textbf{Fig. \ref{fig:DFAvsHKp_1}, middle left}). $|\Delta \hat{H}|$ for the HK method drop below 0.05 for all $H$ with the exception of extreme values (i.e., $H = 0.1, 0.9$; \textbf{Fig. \ref{fig:DFAvsHKp_1}, middle right, respectively}). The same general trend is observed for longer time series ($N = 256, 512, 1024$; \textbf{Figs. \ref{fig:DFAvsHKp_1} \& \ref{fig:DFAvsHKp_2}, middle right, bottom left, and bottom right}). While $|\Delta \hat{H}|$ for DFA drops to reasonable levels for these time series lengths, DFA still tends to be positively biased for $H = 0.1, 0.2, \dots, 0.6$. In contrast, the HK method produces unbiased estimates across the entire range of $H$. In brief, across all $N$ and $H$, the HK method outperforms DFA in that it (i) accurately assesses long-range correlations when the measurement time series is short and (ii) shows minimal dispersion about the central tendency.

Although the DFA estimates $\hat{H}$ with reasonable accuracy for long time series ($Mean$ $|\Delta \hat{H}| \sim 0.05$ for $N > 512$; \textbf{Fig. \ref{fig:DFAvsHKp_2}, bottom left and bottom right}), the HK method estimates $\hat{H}$ with consistently better accuracy than DFA (\textbf{Fig. \ref{fig:DFAvsHKp_2}}). A noteworthy trend is that both methods have a curvilinear error profile, albeit with different forms. The error profile for DFA is concave-up, implying that DFA will be most error-prone when $\hat{H}$ is at both extreme antipersistence and extreme persistence. In contrast, the error profile of the HK method is concave-down, implying that peak error will be in the middle when the time series resembles a wGn. A caveat for that last observation is that as $N \rightarrow 512$, the error in the estimation of $\hat{H}$ using the HK method is smallest as $H \rightarrow 0.1$ and tends to plateau as $H \rightarrow 0.9$ (\textbf{Fig. \ref{fig:DFAvsHKp_2}, bottom left and bottom right}). Thus, practitioners should keep these trends when the estimated $\hat{H}$ values fall within these regions.

\section{Empirical results}

The above results demonstrate the superiority of DFA in estimating the Hurst exponent on synthetic data when underlying dynamics are known. What remains to be learned is the relative performance of the HK method and DFA on human behavioral data. In the following subsections, we present four case studies that demonstrate the superior performance of the HK method in a diverse range of contexts.

\subsection{Context 1: Stride interval time series in a locomotion task}

Healthy and highly adaptable systems---such as the human movement system---display an optimal temporal structure of variability. This ideal structure is described by persistence in stride-to-stride variations indicated by the Hurst exponent, $H$, close to $1$. It implies a temporal structure in consecutive strides that is ordered and stable but also variable and adaptable. DFA has been used in multiple studies to estimate $H$ in stride-to-stride variations in walking \citep{bollens2010does,ducharme2018association,fairley2010effect,hausdorff1995walking,hausdorff1996fractal,jordan2009stability,terrier2011kinematic} and running \citep{agresta2019years,bellenger2019detrended,brahms2020long,fuller2016effect,fuller2017tracking,jordan2006long,lindsay2014effect,nakayama2010variability} under various manipulations of task constraints both on treadmill \citep{agresta2019years,bellenger2019detrended,bollens2010does,ducharme2018association,fairley2010effect,fuller2016effect,fuller2017tracking,hausdorff1995walking,hausdorff1996fractal,jordan2006long,jordan2007speed,jordan2009stability,lindsay2014effect,nakayama2010variability,terrier2011kinematic} and overground \citep{bollens2010does,brahms2020long}. These studies have consistently reported $H$ values close to $1$, at least for young and healthy adults. Furthermore, stride-to-stride variations show a reduction in $H$ in older adults and pathological populations \citep{hausdorff1997altered,hausdorff2001human,hausdorff2009gait,kobsar2014evaluation}. This reduction of persistence in stride-to-stride variations is linked with increased fall risk \citep{hausdorff2001gait,hausdorff2007gait,johansson2016greater,paterson2011stride,toebes2012local}. In short, the Hurst exponent of stride-to-stride variations reflects both the constraints on the movement system due to the task and the physiological health of the movement system. Hence, stride-to-stride variations (e.g., in the stride interval time series) offer an empirical test case to compare downstream performance differences between the HK method and DFA.

\subsubsection{Methods}

Stride interval time series were reanalyzed from a published study on walking and running dynamics on the treadmill and an overground surface \cite{wilson2023multifractality}. Eight adults (5 women and three men; $Mean\pm1s.d.$ age: $30.5\pm11.5$ years) participated in exchange for monetary compensation after providing informed consent approved by the University of Nebraska Medical College's Institutional Review Board. All participants met the following criteria: (i) they could give their informed consent; (ii) they could walk without the aid of a cane or other device; and (iii) they had not been diagnosed with any neurological disease or lower limb disability, injury, or illness.

Participants used a Bodyguard Commercial 312C Treadmill with a top speed of $12.0$ mph and increases/reduction in speed by $0.1$ mph housed in the Balance and Strength Lab at The University of Nebraska at Omaha to do treadmill walking and treadmill running. In addition, participants engaged in overground walking and overground running on the University of Nebraska at Omaha's indoor track, which extends 200 meters and has inner, middle, and outer lanes. Participants donned a Trigno\textsuperscript{TM} 4 Contact FSR (Force Sensitive Resistor) sensor (Delsys Inc., Boston, MA) under each foot. The first and second channels registered relative pressure at the heel and midfoot. A Trigno\textsuperscript{TM} Personal Monitor (TPM) datalogger attached to the participant's body stored the relative pressure data registered FSR sensors.

Participants performed four $20$-min trials across two days. The first day consisted of walking and running either on the treadmill or the indoor track. The second day, separated by at least two but less than seven days, consisted of locomoting on the second surface. On the treadmill locomotion day, two familiarization trials were conducted to estimate the participant’s preferred walking and running speeds based on a previously established protocol \citep{martin1992effects}. Then the participant walked at that speed for $20$ mins. After $5$–$10$ min rest, the participant’s preferred running speed was estimated using the same protocol \citep{martin1992effects}, following which the participant ran at that speed for $20$ mins.

Heel strikes were determined based on the timing associated with the peak pressure of each foot strike from the FSRs. The peak of the $i$th heel strike of the left foot was subtracted from the peak of the $(i-1)$th heel strike of the same foot to determine the stride intervals. The trials produced stride interval time series of various lengths, with the minimum length of $N = 983$. Therefore, all stride interval time series were cropped at $N = 983$ for further analyses. Segments of the original and shuffled stride interval time series of lengths $N = 32, 64, 128, 256, 512, 983$ were submitted to the HK method and DFA. Stride interval time series of all six lengths were shuffled to preserve the probability distribution but destroyed any temporal correlations and submitted to the HK method and DFA. As opposed to the original time series expected to yield $\hat{H} > 0.5$. these shuffled time series were expected to yield an $\hat{H}$ value of $0.5$, indicating an absence of long-range correlations.

We utilized linear mixed-effects (LME) models using Satterthwaite's approximation to examine the effects of locomotion Mode (Walking vs. Running) and Surface (Treadmill vs. Overground) on $\hat{H}$ estimated using the HK method and DFA. Locomotion Mode (Walking vs. Running) and Surface (Treadmill vs. Overground), along with their interactions, served as three fixed effects, and Participant identity served as the random effect (i.e., we allowed the intercept to vary across participants). All mixed-modeling was performed in \texttt{R} \cite{team2013r} using the function \texttt{lmer()} from the package ``nlme" \cite{pinheiro2007linear} and the function \texttt{anova()} from the package ``lmertest" \cite{kuznetsova2015package}. Statistical significance was set at the Type I error rate of $5\%$.

\subsubsection{Results}

\begin{figure*}
\centering
\includegraphics[width=0.75\textwidth]{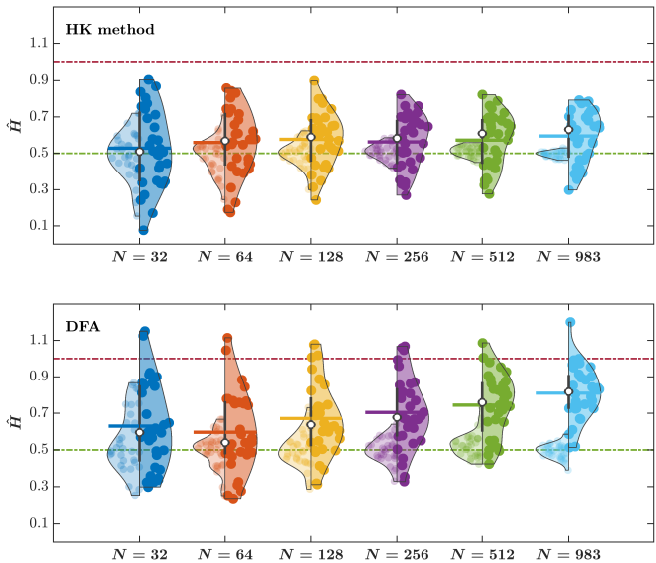}
\caption{\textbf{The Hurst exponent, \boldmath$\hat{H}$, for stride interval time series estimated using the HK method do not depend on the time series length \boldmath$N$, but \boldmath$\hat{H}$ estimated using DFA show a strong dependence on \boldmath$N$, resulting in larger \boldmath$\hat{H}$ for larger \boldmath$N$.} The right and the left violin plots represent the distribution of $\hat{H}$ for the original and shuffled stride interval time series, respectively, estimated using the HK method (top) and DFA (bottom). Vertical lines represent the interquartile range of the original $\hat{H}$ values, white circles represent the median value of $\hat{H}$, and horizontal lines represent the $Mean$ value of $\hat{H}$ for the original stride interval time series. Horizontal dash-dotted green and red lines indicate $\hat{H} = 0.5$ and $\hat{H} = 1$, respectively.}
\label{fig:DFAvsHKp_3}
\end{figure*}

The central tendencies—$Mean$ and $Median$---of $\hat{H}$ for stride interval time series estimated using the HK method, as well as the distribution of $\hat{H}$, do not depend on the time series length $N$, except for $N = 32$ for which the HK method yields marginally smaller $\hat{H}$ (\textbf{Fig. \ref{fig:DFAvsHKp_3}, top}). In contrast, while the $Mean$ and $Median$ $\hat{H}$ for stride interval time series estimated using DFA do not appear to differ between $N = 32$ and $N = 64$, they show a consistent and linear increase with $N$ after that. Furthermore, while the $\hat{H}$ values estimated using the HK method lie within the tight bounds of $[0,1]$, the $\hat{H}$ values estimated using DFA often exceed the upper bound of $1$  (\textbf{Fig. \ref{fig:DFAvsHKp_3}, bottom}). Another notable distinction is a narrower range of $\hat{H}$ for the shuffled stride interval time series estimated using the HK method compared to DFA. Overall, the HK method estimates $\hat{H}$ that show smaller dispersion about the central tendency and lesser dependence on the length of the stride interval time series.

To investigate the sensitivity of both methods to task constraints, we analyzed the influence of locomotion Mode and Surface on $\hat{H}$ values estimated using both methods. We submitted the $\hat{H}$ values estimated using both methods to linear mixed-effects modeling with Satterthwaite's approximation for finite sample size \cite{luke2017evaluating}. We performed this modeling separately for each time series length $N = 32, 64, 128, 256, 512, 1024$. \textbf{Tables 1 \& 2} describe the model outcomes.

\newpage

\noindent{}\textbf{Table 1.} Outcomes of linear mixed-effects modeling with Satterthwaite's approximation for small sample size, examining the influence of locomotion Mode and Surface on the Hurst exponent, $\hat{H}$, estimated using the HK method for stride interval time series of length $N = 32, 64, 128, 256, 512, 983$.

\noindent
\begin{tabular}{|l||*{5}{c|}}\hline
&\makebox[3em]{\textbf{Mean Sq}}&\makebox[3em]{\textbf{Sum Sq}}&\makebox[3em]{\textbf{DF}}
&\makebox[3em]{\textbf{\textit{F}}}&\makebox[3em]{\textbf{\textit{P}}\boldmath$^{*}$}\\\hline\hline
\boldmath$N = 32$ &&&&&\\\hline
Mode &0.199&0.199&1,32&8.277&\textbf{0.007}\\\hline
Surface &0.346&0.346&1,32&14.397&\boldmath$<\;$\textbf{0.001}\\\hline
Mode $\times$ Surface &0.105&0.105&1,32&4.388&\textbf{0.044}\\\hline
\boldmath$N = 64$ &&&&&\\\hline
Mode &0.217&0.217&1,24&12.662&\textbf{0.002}\\\hline
Surface &0.174&0.174&1,24&10.132&\textbf{0.004}\\\hline
Mode $\times$ Surface &0.028&0.028&1,24&1.642&0.212\\\hline
\boldmath$N = 128$ &&&&&\\\hline
Mode &0.187&0.187&1,32&15.056&\boldmath$<\;$\textbf{0.001}\\\hline
Surface &0.124&0.124&1,32&10.004&\textbf{0.003}\\\hline
Mode $\times$ Surface &0.022&0.022&1,32&1.789&0.191\\\hline
\boldmath$N = 256$ &&&&&\\\hline
Mode &0.127&0.127&1,24&9.389&\textbf{0.005}\\\hline
Surface &0.112&0.112&1,24&8.229&\textbf{0.008}\\\hline
Mode $\times$ Surface &0.003&0.003&1,24&0.227&0.638\\\hline
\boldmath$N = 512$ &&&&&\\\hline
Mode &0.152&0.152&1,24&13.093&\textbf{0.001}\\\hline
Surface &0.150&0.150&1,24&12.930&\textbf{0.001}\\\hline
Mode $\times$ Surface &0.000&0.000&1,24&0.043&0.838\\\hline
\boldmath$N = 983$ &&&&&\\\hline
Mode &0.128&0.128&1,24&14.794&\boldmath$<\;$\textbf{0.001}\\\hline
Surface &0.119&0.119&1,24&13.759&\textbf{0.001}\\\hline
Mode $\times$ Surface &0.000&0.000&1,24&0.097&0.758\\\hline
\end{tabular}
\noindent{}\\
\noindent{}$^{*}$Boldfaced values indicate significant differences at the two-tailed alpha of 0.05.

\noindent{}\\

\newpage

\noindent{}\textbf{Table 2.} Outcomes of linear mixed-effects modeling with Satterthwaite's approximation for small sample size, examining the influence of locomotion Mode and Surface on the Hurst exponent, $\hat{H}$, estimated using DFA for stride interval time series of length $N = 32, 64, 128, 256, 512, 983$.

\noindent
\begin{tabular}{|l||*{5}{c|}}\hline
&\makebox[3em]{\textbf{Mean Sq}}&\makebox[3em]{\textbf{Sum Sq}}&\makebox[3em]{\textbf{DF}}
&\makebox[3em]{\textbf{\textit{F}}}&\makebox[3em]{\textbf{\textit{P}}\boldmath$^{*}$}\\\hline\hline
\boldmath$N = 32$ &&&&&\\\hline
Mode &0.144&0.144&1,32&3.960&0.055\\\hline
Surface &0.276&0.276&1,32&7.584&\textbf{0.009}\\\hline
Mode $\times$ Surface &0.148&0.148&1,32&4.053&0.053\\\hline
\boldmath$N = 64$ &&&&&\\\hline
Mode &0.284&0.284&1,24&11.346&\textbf{0.003}\\\hline
Surface &0.254&0.254&1,24&10.140&\textbf{0.004}\\\hline
Mode $\times$ Surface &0.111&0.111&1,24&4.447&\textbf{0.046}\\\hline
\boldmath$N = 128$ &&&&&\\\hline
Mode &0.200&0.200&1,24&9.280&\textbf{0.006}\\\hline
Surface &0.159&0.159&1,24&7.385&\textbf{0.012}\\\hline
Mode $\times$ Surface &0.091&0.091&1,24&4.230&0.051\\\hline
\boldmath$N = 256$ &&&&&\\\hline
Mode &0.060&0.060&1,24&1.958&0.174\\\hline
Surface &0.127&0.127&1,24&4.129&0.053\\\hline
Mode $\times$ Surface &0.002&0.002&1,24&0.077&0.784\\\hline
\boldmath$N = 512$ &&&&&\\\hline
Mode &0.148&0.148&1,32&6.754&\textbf{0.014}\\\hline
Surface &0.076&0.076&1,32&3.442&0.073\\\hline
Mode $\times$ Surface &0.017&0.017&1,32&0.758&0.391\\\hline
\boldmath$N = 983$ &&&&&\\\hline
Mode &0.161&0.161&1,24&12.951&\textbf{0.001}\\\hline
Surface &0.062&0.062&1,24&4.960&\textbf{0.036}\\\hline
Mode $\times$ Surface &0.009&0.009&1,24&0.714&0.406\\\hline
\end{tabular}
\noindent{}\\
\noindent{}$^{*}$Boldfaced values indicate significant differences at the two-tailed alpha of 0.05.

\noindent{}\\

\begin{figure*}
\centering
\includegraphics[width=0.75\textwidth]{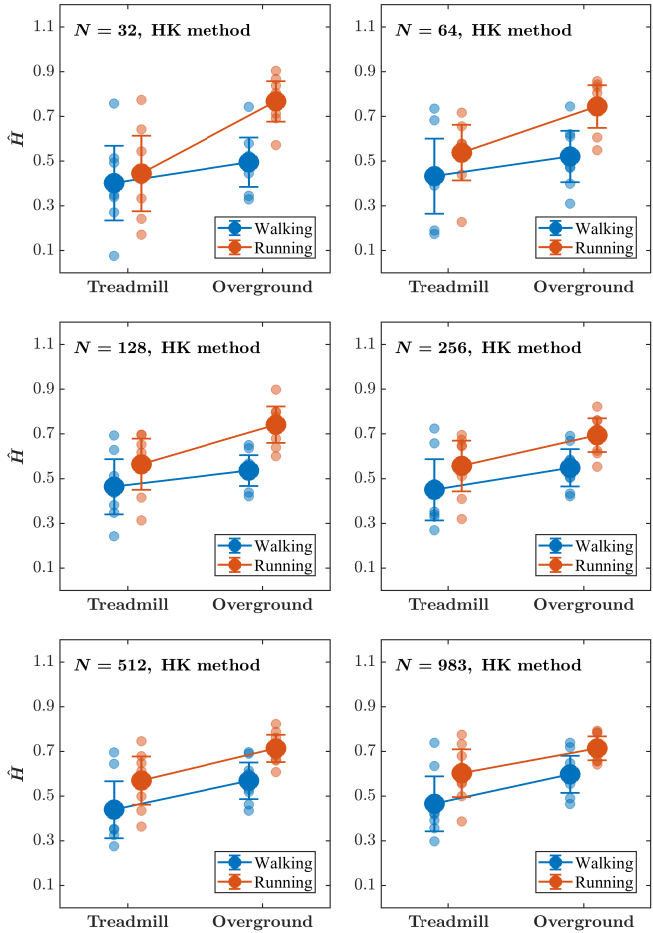}
\caption{\textbf{The effects of locomotion mode and surface on the Hurst exponent, \boldmath$\hat{H}$, estimated using the HK method do not depend o such as uniform or truncated Gaussian distributions, etc. the stride interval time series length (see Table 1 for the outcomes of the statistical tests).} Each panel plots the $Mean$ values of $\hat{H}$, estimated using the HK method for stride interval time series of length $N = 32, 64, 128, 256, 512, 983$. Light blue and light red circles indicate $\hat{H}$ values for individual participants in the respective conditions. Error bars indicate $95$\% CI across $8$ participants.}
\label{fig:DFAvsHKp_4}
\end{figure*}

Linear-mixed effects modeling of $\hat{H}$ estimated using the HK method revealed that Running is associated with greater $\hat{H}$ (i.e., stronger long-range correlations in stride-to-stride variations) compared to Walking, and Overground locomotion is associated with greater $\hat{H}$ compared to Treadmill locomotion (\textbf{Fig. \ref{fig:DFAvsHKp_4}}; \textbf{Table 1}). These results are supported by previous studies that have reported similar effects of locomotion Mode and Surface on the long-range correlations in stride-to-stride fluctuations \cite{bollens2010does,wilson2023multifractality}. These results also remain consistent across all values of $N$ ($32, 64, 128, 256, 512, 1024$), suggesting that the HK method is sensitive to task constraints for stride interval time series as short as 32 strides. Lastly, it is noteworthy from a movement science perspective that locomotion Mode and Surface exert their influence independently. However, this effect must be replicated, given the relatively small sample size.

\begin{figure*}
\centering
\includegraphics[width=0.75\textwidth]{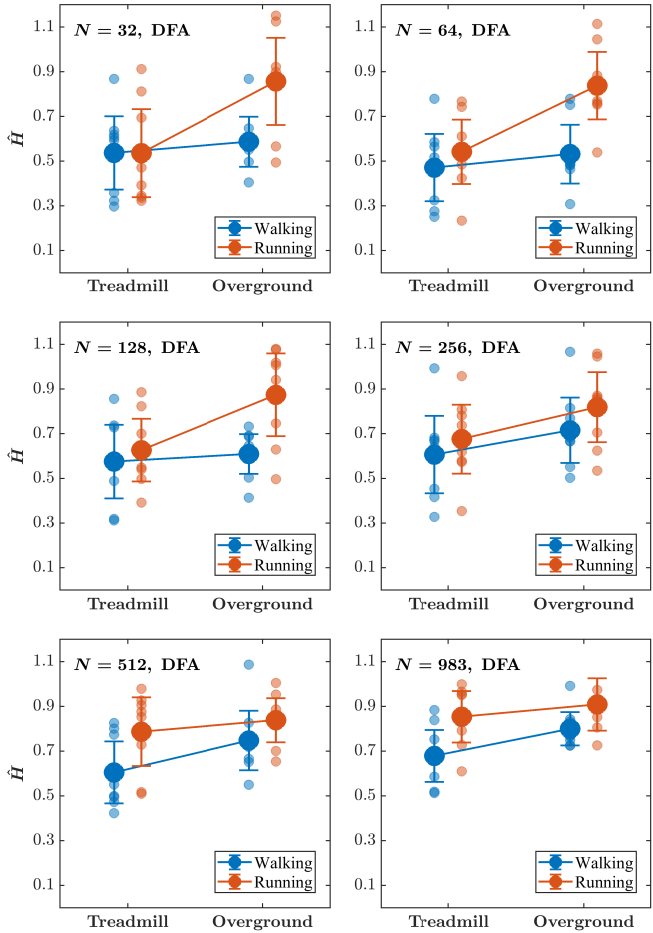}
\caption{\textbf{The effects of locomotion mode and surface on the Hurst exponent, \boldmath$\hat{H}$, estimated using DFA wax and wane depending on the stride interval time series length (see Table 2 for the outcomes of the statistical tests).} Each panel plots the $Mean$ values of $\hat{H}$, estimated using DFA for stride interval time series of length $N = 32, 64, 128, 256, 512, 983$. Light blue and light red circles indicate $\hat{H}$ values for individual participants in the respective conditions. Error bars indicate $95$\% CI across $8$ participants.}
\label{fig:DFAvsHKp_5}
\end{figure*}

In contrast to the HK method, the results for the linear-mixed effects modeling of $\hat{H}$ estimated using DFA wax and wane depending on the stride interval time series length (\textbf{Fig. \ref{fig:DFAvsHKp_5}}; \textbf{Table 2}). For $N = 32$, Overground locomotion seems to produce greater $\hat{H}$ values than Treadmill locomotion. However, for $N = 64$, Running is associated with greater $\hat{H}$ than Walking, and the interaction effect of locomotion Mode and Surface appears. Both factors show an effect for $N = 128$, but then the effects of both factors disappear for $N = 256$. Then again, for $N = 512$---the typical recommendation for the application of DFA in gait analysis \cite{kuznetsov2017power}, Running is associated with greater $\hat{H}$ compared to Walking, and for $N = 983$, the effect of locomotion surface meets conventional levels of statistical significance.

These results suggest that DFA, when used with short empirical time series, increases the likelihood of the Type II error because we failed to find consistent effects that are present when the time series is long. Therefore, DFA should be reasonably accurate based on our simulations. This is problematic from the perspective of accumulating knowledge in movement science (and other fields) because such findings do not meet conventional levels of statistical significance still pervasive in scientific literature irrespective of relentless criticism \cite{amrhein2019scientists,berger1987testing,halsey2019reign,ioannidis2019have,lieberman2009type}. We argue that this phenomenon may be more prevalent than previously thought in studies using the Hurst exponent as a dependent variable, which could have severe theoretical consequences. The above-described results on stride interval time series strongly support the idea that adopting the HK method in favor of DFA could drastically reduce the likelihood of Type II errors in behavioral sciences where $H$ is a critical dependent variable. On a more substantive level, we recognize that the HK method produces lower $H$ values than are typically observed in the gait literature. Whether these specific results generalize to other contexts is a matter of extensive replication.

\subsection{Context 2: Intertap interval time series in a syncopation task}

It has now been well established that the series of time intervals produced in repetitive tapping also show persistence or long-range correlations in sample-to-sample variations \cite{delignieres2009long,lemoine2006testing,torre2008unraveling,torre2011long}. Instead of being a universally prevalent generic property of sensory time series, these long-range correlations in taping interval time series constitute a constant and recognizable characteristic of individuals performing a specific tapping activity, e.g., synchronizing with pacing signals of different fractal properties \cite{coey2015complexities,delignieres2009long,delignieres2014strong,stephen2008strong,torre2011long}. Tapping interval time series thus offer another empirical test case to compare the performance of the HK method and DFA.

\subsubsection{Methods}

Intertap interval time series were collected. Intertap intervals were recorded as participants pressed the letter ``M” on their keyboard at a pace they could maintain for $1$ minute. Participants performed tapping for $8$ min in four conditions: three paced conditions: ``Persistent,” ``Random,” and ``Periodic,” and one without pacing,” ``Free.” In paced conditions, participants synchronized their finger taps to the metronome by pressing the letter ``M.” In the Persistent condition, participants synchronized their tapping to a variable and structured metronome with interbeat interval time series exhibiting a Hurst’s exponent, $H$, of $1.0$. In the Random condition, participants synchronized their tapping to a non-correlated metronome with interbeat interval time series exhibiting $H$ of $0.5$. In the Periodic condition, participants synchronized their taps to an invariant metronome (i.e., traditional metronome). Finally, in the Free condition, participants pressed the letter ``M” at a self-selected pace. The $Mean$ and $SD$s of Persistent and Random signals were set equal to each participant's preferred tapping characteristics. The Periodic signal period was set equal to each Participant's preferred tapping interval. The order of the four conditions was randomized for each participant.

The tapping trial was successful if the number of taps in the pacing condition was within $10$\% of the self-paced condition. Nineteen participants who fulfilled this criterion in all three pacing conditions were included for further analysis. Intertap interval time series of lengths $N = 32, 64, 128, 256$ were submitted to the HK method and DFA. Intertap interval time series of all four lengths were shuffled to preserve the probability distribution but destroyed any temporal correlations and submitted to the HK method and DFA. As opposed to the original time series expected to yield $\hat{H} > 0.5$. these shuffled time series were expected to yield an $\hat{H}$ value of $0.5$, indicating an absence of long-range correlations.

We utilized LME models using Satterthwaite's approximation to examine the effects of the Pacing condition on $\hat{H}$ values for the tapping interval time series estimated using the HK method and DFA. Pacing condition served as the fixed effect, and Participant identity was included as a random effect. All mixed-modeling was performed in \texttt{R} \cite{team2013r} using the function \texttt{lmer()} from the package ``nlme" \cite{pinheiro2007linear} and the function \texttt{anova()} from the package ``lmertest" \cite{kuznetsova2015package}. Statistical significance was set at the Type I error rate of $5\%$.

\subsubsection{Results}

\begin{figure*}
\centering
\includegraphics[width=0.75\textwidth]{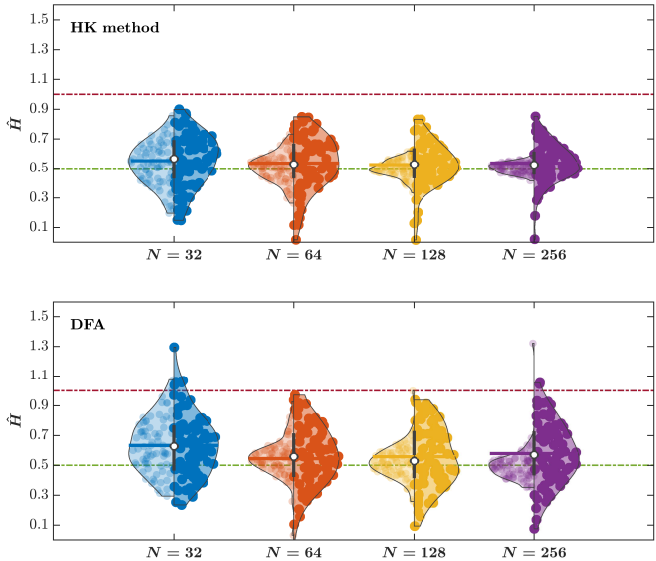}
\caption{\textbf{The Hurst exponent, \boldmath$\hat{H}$, for the finger tapping interval time series estimated using the HK method do not depend on the time series length \boldmath$N$, but \boldmath$\hat{H}$ estimated using DFA show a strong dependence on \boldmath$N$, resulting in larger \boldmath$\hat{H}$ for smaller and larger \boldmath$N$.} The right and the left violin plots represent the distribution of $\hat{H}$ for the original and shuffled tapping interval time series, respectively, estimated using the HK method (top) and DFA (bottom). Vertical lines represent the interquartile range of the original $\hat{H}$ values, white circles represent the median value of $\hat{H}$, and horizontal lines represent the $Mean$ value of $\hat{H}$ for the original stride interval time series. Horizontal dash-dotted green and red lines indicate $\hat{H} = 0.5$ and $\hat{H} = 1$, respectively.}
\label{fig:DFAvsHKp_6}
\end{figure*}

The central tendencies—$Mean$ and $Median$---of $\hat{H}$ for the tapping interval time series estimated using the HK method, as well as the distribution of $\hat{H}$, do not depend on the time series length $N$, except for $N = 32$ for which the HK method yields marginally larger $\hat{H}$ (\textbf{Fig. \ref{fig:DFAvsHKp_6}, top}). In contrast, while the $Mean$ and $Median$ $\hat{H}$ for tapping interval time series estimated using DFA do not appear to depend on the time series length $N$, the $\hat{H}$ values show a larger dispersion around the $Mean$ compared to the counterparts estimates using the HK method (\textbf{Fig. \ref{fig:DFAvsHKp_6}, bottom}). While the $\hat{H}$ values estimated using the HK method lie with the tight bounds of $[0,1]$, the $\hat{H}$ values estimated using DFA often exceed the upper bound of $1$. Another notable distinction is a narrower range of $\hat{H}$ for the shuffled tapping interval time series estimated using the HK method compared to DFA. Overall, $\hat{H}$ of tapping interval time series estimated using the HK method estimates show smaller dispersion about the central tendency and are more consistent with the theory of the Hurst exponent.

To investigate the sensitivity of both methods to task constraints, we analyzed the influence of the Pacing condition on $\hat{H}$ values for the tapping interval estimated using both methods. We performed this modeling separately for each time series length $N = 32, 64, 128, 256$. \textbf{Tables 3 \& 4} describe the model outcomes.

\newpage

\noindent{}\textbf{Table 3.} Outcomes of linear mixed-effects modeling with Satterthwaite's approximation for small sample size, examining the influence of the Pacing condition on the Hurst exponent, $\hat{H}$, estimated using the HK method for the tapping interval time series of length $N = 32, 64, 128, 256$.

\noindent
\begin{tabular}{|l||*{5}{c|}}\hline
&\makebox[3em]{\textbf{Mean Sq}}&\makebox[3em]{\textbf{Sum Sq}}&\makebox[3em]{\textbf{DF}}
&\makebox[3em]{\textbf{\textit{F}}}&\makebox[3em]{\textbf{\textit{P}}\boldmath$^{*}$}\\\hline\hline
\boldmath$N = 32$ &&&&&\\\hline
Pacing condition &0.170&0.057&3,57&1.8591&0.147\\\hline
\boldmath$N = 64$ &&&&&\\\hline
Pacing condition &0.763&0.254&3,76&10.727&\boldmath$<\;$\textbf{0.001}\\\hline
\boldmath$N = 128$ &&&&&\\\hline
Pacing condition &0.763&0.254&3,76&19.113&\boldmath$<\;$\textbf{0.001}\\\hline
\boldmath$N = 256$ &&&&&\\\hline
Pacing condition &0.532&0.177&3,57&14.787&\boldmath$<\;$\textbf{0.001}\\\hline
\end{tabular}
\noindent{}\\
\noindent{}$^{*}$Boldfaced values indicate significant differences at the two-tailed alpha of 0.05.

\noindent{}\\

\noindent{}\textbf{Table 4.} Outcomes of linear mixed-effects modeling with Satterthwaite's approximation for small sample size, examining the influence of the Pacing condition on the Hurst exponent, $\hat{H}$, estimated using DFA for the tapping interval time series of length $N = 32, 64, 128, 256$.

\noindent
\begin{tabular}{|l||*{5}{c|}}\hline
&\makebox[3em]{\textbf{Mean Sq}}&\makebox[3em]{\textbf{Sum Sq}}&\makebox[3em]{\textbf{DF}}
&\makebox[3em]{\textbf{\textit{F}}}&\makebox[3em]{\textbf{\textit{P}}\boldmath$^{*}$}\\\hline\hline
\boldmath$N = 32$ &&&&&\\\hline
Pacing condition &0.763&0.254&3,76&10.727&\boldmath$<\;$\textbf{0.001}\\\hline
\boldmath$N = 64$ &&&&&\\\hline
Pacing condition &0.519&0.173&3,76&2.567&0.061\\\hline
\boldmath$N = 128$ &&&&&\\\hline
Pacing condition &1.355&0.451&3,76&18.774&\boldmath$<\;$\textbf{0.001}\\\hline
\boldmath$N = 256$ &&&&&\\\hline
Pacing condition &1.560&0.520&3,76&22.876&\boldmath$<\;$\textbf{0.001}\\\hline
\end{tabular}
\noindent{}\\
\noindent{}$^{*}$Boldfaced values indicate significant differences at the two-tailed alpha of 0.05.

\noindent{}\\

\begin{figure*}
\centering
\includegraphics[width=0.75\textwidth]{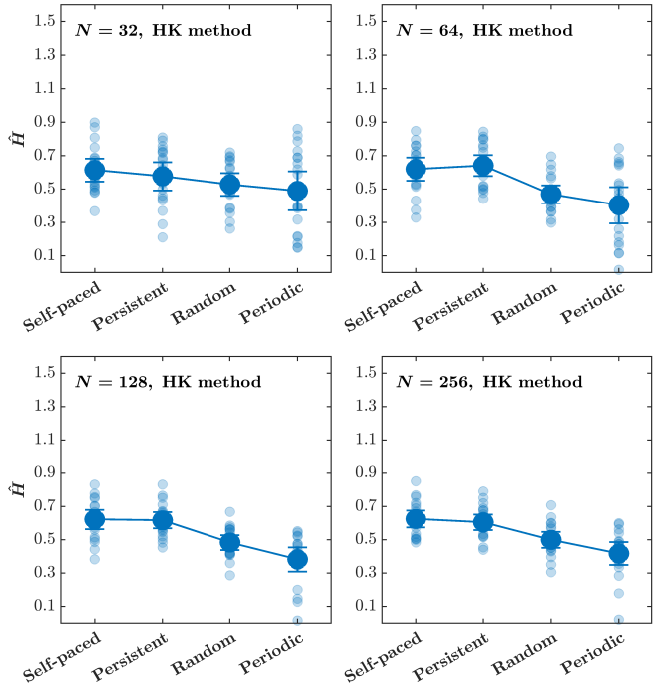}
\caption{\textbf{The effects of pacing conditions on the Hurst exponent, \boldmath$\hat{H}$, estimated using the HK method do not depend on the tapping interval time series length (see Table 2 for the outcomes of the statistical tests).} Each panel plots the $Mean$ values of $\hat{H}$, estimated using the HK method for the tapping interval time series of length $N = 32, 64, 128, 256, 512, 983$. Light blue circles indicate $\hat{H}$ values for individual participants. Error bars indicate $95$\% CI across $19$ participants.}
\label{fig:DFAvsHKp_7}
\end{figure*}

\begin{figure*}
\centering
\includegraphics[width=0.75\textwidth]{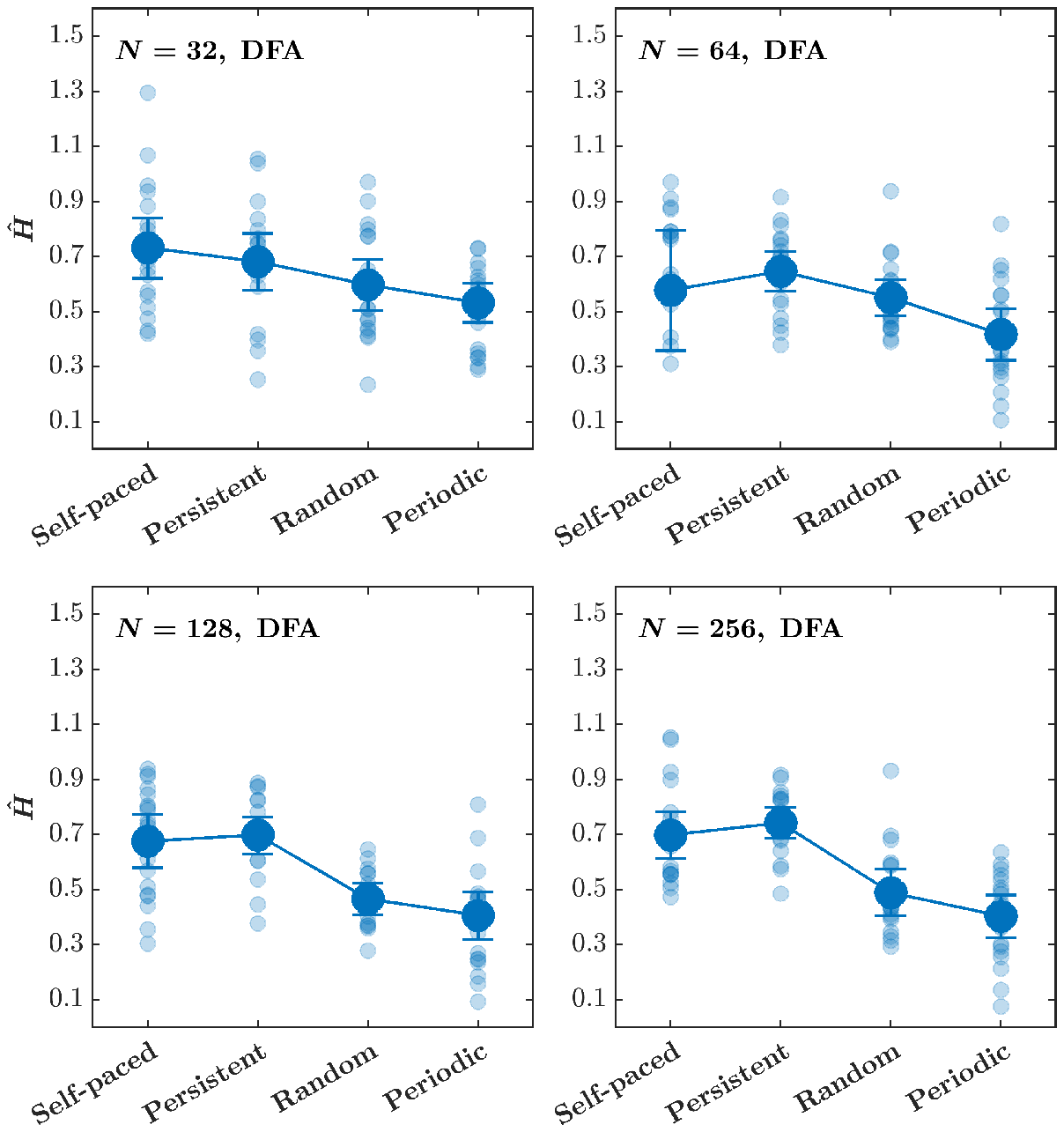}
\caption{\textbf{The effects of pacing condition on the Hurst exponent, \boldmath$\hat{H}$, estimated using DFA wax and wane depending on the tapping interval time series length (see Table 4 for the outcomes of the statistical tests).} Each panel plots the $Mean$ values of $\hat{H}$, estimated using DFA for the tapping interval time series of length $N = 32, 64, 128, 256, 512, 983$. Light blue circles indicate $\hat{H}$ values for individual participants. Error bars indicate $95$\% CI across $19$ participants.}
\label{fig:DFAvsHKp_8}
\end{figure*}

The $\hat{H}$ values estimated using the HK method differed across the Pacing conditions for the tapping interval time series of length $N = 64, 128, 256$ but not for $N = 32$ (\textbf{Fig. \ref{fig:DFAvsHKp_7}}; \textbf{Table 3}). In other words, the $\hat{H}$ values estimated using the HK method are sensitive to the pacing condition for the tapping interval time series comprising at least $64$ intervals, and this sensitivity is consistent across progressively longer time series. In contrast, the effect of the Pacing condition on the $\hat{H}$ values estimated using DFA wax and wane depending on the tapping interval time series length, appearing for $N = 32$ but disappearing for $N = 64$ (\textbf{Fig. \ref{fig:DFAvsHKp_8}}; \textbf{Table 4}). Hence, the HK method yields more consistent effects of the different temporal structures of the pacing signal on the Hurst exponent of tapping intervals in a syncopation task. These results align with the above results on stride interval time series and strengthen the argument that DFA increases the likelihood of Type II error and makes an even more compelling case for adopting the HK method over the age-old DFA for estimating the Hurst exponent in behavioral sciences.

\subsection{Context 3: Reaction time (RT) time series in simple and choice RT tasks}

Reaction time (RT) is a workhorse of cognitive science and many other areas of psychological science. Often, RTs are collected from many (sometimes hundreds or thousands) of trials under the assumption that for a given experimental task, there exists a ``true” RT that can be extracted from repeated sampling. The key to that assumption is that variation in RTs reflects independent white noise. However, persistence in sample-to-sample variations or $0.5 < H < 1$ is not limited to predominantly movement tasks---such as walking, running, or finger tapping, but tasks involving spatial or temporal interval estimation also seem to show $1/f^{\alpha}$ noise unambiguously \cite{holden2009dispersion,kuznetsov2011effects,wagenmakers2004estimation}. Simple RT tasks and choice RT tasks, such as lexical decision-making, also seem to provide unambiguous information about the state of the physiological system in that the RT time series in these tasks also yields $0.5 < H < 1$ \cite{gilden1995noise,irrmischer2018negative,malone2014dynamic,van2003self,wagenmakers2004estimation}. Therefore, we also compare the performance of the HK method and DFA using RTs from three tasks (a simple RT, a forced-choice RT, and time estimation task), all conducted in a similar experimental format.

\subsubsection{Methods}

RT time series were reanalyzed from a published study  \cite{wagenmakers2004estimation}. Six healthy adults responded to the Arabic digits $1, 2, 3, 4, 6, 7, 8, \mathrm{and}\;9$ displayed on a computer screen. The experimental phase consisted of $1024$ stimuli after $24$ practice stimuli, with each stimulus appearing equally frequently in a randomized order for each task and participant. Each participant completed three tasks: (i) simple RT, in which they pressed the ``?/” key with their right index finger immediately after detecting the stimulus. In addition to instructing participants to avoid anticipations, feedback ``TOO FAST” was presented for two seconds following responses $< 100$ ms to prevent anticipatory responding (e.g., \cite{snodgrass1967some}). (ii) Choice RT, in which they pressed the ``?/” key with their right index finger in response to an even number and pressed the ``z” key in response to an odd number, ``as fast as possible without making errors.” (iii) one second time interval estimation, in which participants pressed the ``?/” key with their right index finger to mark an estimated time interval of one second after each stimulus was presented. The task order was counterbalanced across participants.

Each task (i.e., Simple RT, Choice RT, and time interval Generation) was performed with a relatively short response-stimulus interval (RSI) and a Long RSI, yielding a total of 3 tasks $\times$ $2$ RSI $=$ six sessions conducted on different days. One set of RSIs was randomly drawn from a uniform distribution that extended from $200$ ms to $600$ ms, and the set of long RSIs was obtained by adding a constant $600$ ms to the set of short RSIs, and hence the long RSIs varied between $800$ ms and $1200$ ms. The order of the RSIs was randomized for each task and participant. The experiment---aimed at collecting RT time series of length $N = 1024$---yielded RT time series with a minimum length $N = 1020$. RT time series of lengths $N = 32, 64, 128, 256, 512, 1020$ were submitted to the HK method and DFA. RT time series of all four lengths were shuffled to preserve the probability distribution but destroyed any temporal correlations and submitted to the HK method and DFA. As opposed to the original time series expected to yield $\hat{H} > 0.5$. these shuffled time series were expected to yield an $\hat{H}$ value of $0.5$, indicating an absence of long-range correlations.

We utilized LME models using Satterthwaite's approximation to examine the effects of Task and RSI on $\hat{H}$ values estimated using both methods. Pacing condition served as the fixed effect, and Participant identity served as the random effect. Task (Simple RT vs. Choice RT vs. Generation) and RSI (Short vs. Long), along with their interactions, served as three fixed effects, and Participant identity served as the random effect. All mixed-modeling was performed in \texttt{R} \cite{team2013r} using the function \texttt{lmer()} from the package ``nlme" \cite{pinheiro2007linear} and the function \texttt{anova()} from the package ``lmertest" \cite{kuznetsova2015package}. Statistical significance was set at the Type I error rate of $5\%$.

\subsubsection{Results}

\begin{figure*}
\centering
\includegraphics[width=0.75\textwidth]{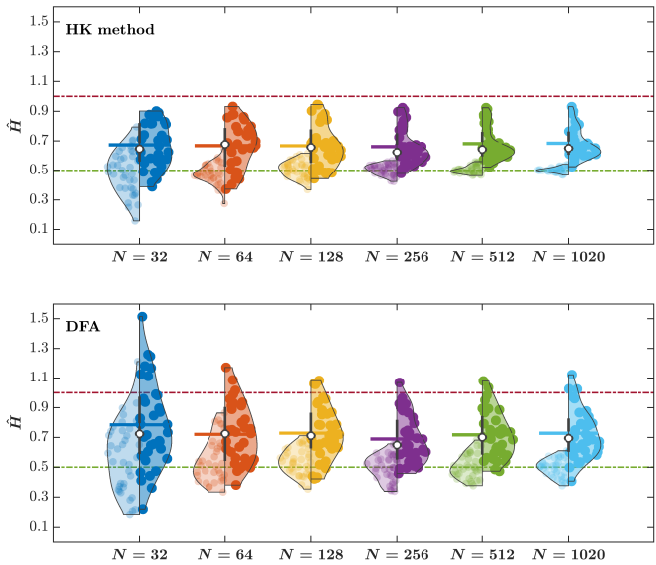}
\caption{\textbf{The Hurst exponent, \boldmath$\hat{H}$, for the response-stimulus interval time series estimated using the HK method do not depend on the time series length \boldmath$N$, but \boldmath$\hat{H}$ estimated using DFA show a strong dependence on \boldmath$N$, resulting in larger \boldmath$\hat{H}$ for smaller and larger \boldmath$N$.} The right and the left violin plots represent the distribution of $\hat{H}$ for the original and shuffled response-stimulus interval time series, respectively, estimated using the HK method (top) and DFA (bottom). Vertical lines represent the interquartile range of the original $\hat{H}$ values, white circles represent the median value of $\hat{H}$, and horizontal lines represent the $Mean$ value of $\hat{H}$ for the original stride interval time series. Horizontal dash-dotted green and red lines indicate $\hat{H} = 0.5$ and $\hat{H} = 1$, respectively.}
\label{fig:DFAvsHKp_9}
\end{figure*}

The central tendencies—$Mean$ and $Median$---of $\hat{H}$ for RT time series estimated using the HK method, as well as the distribution of $\hat{H}$, do not depend on the time series length $N$, showing only marginal dependence of the distribution of $\hat{H}$ on $N$ (\textbf{Fig. \ref{fig:DFAvsHKp_9}, top}). In contrast, while the $Mean$ and $Median$ $\hat{H}$ for the RT time series estimated using DFA do not differ among $N = 64, 128, 512, 1024$, the values are visibly greater for $N = 32$ and smaller for $N = 256$ (\textbf{Fig. \ref{fig:DFAvsHKp_9}, bottom}). Furthermore, while the $\hat{H}$ values estimated using the HK method lie with the tight bounds of $[0,1]$, the $\hat{H}$ values estimated using the DFA frequently exceed the upper bound of $1$, especially for short time series. Another notable distinction is a narrower range of $\hat{H}$ for the shuffled RT time series estimated using the HK method compared to the DFA. Overall, and similar to the results above, the HK method estimates $\hat{H}$ that show smaller dispersion about the central tendency and lesser dependence on the length of the RT time series.

To investigate the sensitivity of both methods to task constraints, we analyzed the influence of the Task and RSI on $\hat{H}$ on $\hat{H}$ values for the tapping interval estimated using both methods. We performed this modeling separately for each time series length $N = 32, 64, 128, 256$. \textbf{Tables 5 \& 6} describe the model outcomes.

Time interval generation is associated with greater $\hat{H}$ (i.e., stronger long-range correlations in sample-to-sample variations) compared to simple RT and choice RT, and RSI shows significant interaction with the task (\textbf{Fig. \ref{fig:DFAvsHKp_12}}). Furthermore, these effects of Task and Task $\times$ RSI interaction remain consistent across all values of $N$ ($32, 64, 128, 256, 512, 1020$; \textbf{Table 5}), suggesting that the HK method is sensitive to task constraints for RT time series as short as 32 RTs. This result dovetails with what we found in the stride interval time series and tapping interval time series reported above.

\noindent{}\\ \noindent{}\\

\newpage

\noindent{}\textbf{Table 5.} Outcomes of linear mixed-effects modeling with Satterthwaite's approximation for small sample size, examining the influence of Task and RSI on the Hurst exponent, $\hat{H}$, estimated using the HK method for the RT time series of length $N = 32, 64, 128, 256, 512, 983$.

\noindent
\begin{tabular}{|l||*{5}{c|}}\hline
&\makebox[3em]{\textbf{Mean Sq}}&\makebox[3em]{\textbf{Sum Sq}}&\makebox[3em]{\textbf{DF}}
&\makebox[3em]{\textbf{\textit{F}}}&\makebox[3em]{\textbf{\textit{P}}\boldmath$^{*}$}\\\hline\hline
\boldmath$N = 32$ &&&&&\\\hline
Task &0.228&0.114&2,36&11.634&\boldmath$< 0.001$\\\hline
RSI &0.031&0.031&1,36&3.202&0.082\\\hline
Task $\times$ RSI &0.232&0.116&2,36&11.840&\boldmath$< 0.001$\\\hline
\boldmath$N = 64$ &&&&&\\\hline
Task &0.226&0.113&2,36&11.913&\boldmath$< 0.001$\\\hline
RSI &0.005&0.005&1,36&0.554&0.462\\\hline
Task $\times$ RSI &0.257&0.128&2,36&13.501&\boldmath$< 0.001$\\\hline
\boldmath$N = 128$ &&&&&\\\hline
Task &0.145&0.073&2,36&9.826&\boldmath$< 0.001$\\\hline
RSI &0.022&0.022&1,36&3.039&0.090\\\hline
Task $\times$ RSI &0.318&0.159&2,36&21.538&\boldmath$< 0.001$\\\hline
\boldmath$N = 256$ &&&&&\\\hline
Task &0.061&0.030&2,36&5.985&\textbf{0.006}\\\hline
RSI &0.012&0.012&1,36&2.342&0.135\\\hline
Task $\times$ RSI &0.258&0.129&2,36&25.337&\boldmath$< 0.001$\\\hline
\boldmath$N = 512$ &&&&&\\\hline
Task &0.088&0.044&2,30&10.154&\boldmath$< 0.001$\\\hline
RSI &0.002&0.002&1,30&0.470&0.498\\\hline
Task $\times$ &0.181&0.091&2,30&20.990&\boldmath$< 0.001$\\\hline
\boldmath$N = 1020$ &&&&&\\\hline
Task &0.088&0.044&2,36&14.371&\boldmath$< 0.001$\\\hline
RSI &0.001&0.001&1,36&0.414&0.5241\\\hline
Task $\times$ RSI &0.164&0.082&2,36&26.747&\boldmath$< 0.001$\\\hline
\end{tabular}
\noindent{}\\
\noindent{}$^{*}$Boldfaced values indicate significant differences at the two-tailed alpha of 0.05.

\noindent{}\\

\newpage

\noindent{}\textbf{Table 6.} Outcomes of linear mixed-effects modeling with Satterthwaite's approximation for small sample size, examining the influence of Task and RSI on the Hurst exponent, $\hat{H}$, estimated using DFA for the RT time series of length $N = 32, 64, 128, 256, 512, 983$.

\noindent
\begin{tabular}{|l||*{5}{c|}}\hline
&\makebox[3em]{\textbf{Mean Sq}}&\makebox[3em]{\textbf{Sum Sq}}&\makebox[3em]{\textbf{DF}}
&\makebox[3em]{\textbf{\textit{F}}}&\makebox[3em]{\textbf{\textit{P}}\boldmath$^{*}$}\\\hline\hline
\boldmath$N = 32$ &&&&&\\\hline
Task &0.282&0.141&2,36&2.287&0.116\\\hline
RSI &0.028&0.028&1,36&0.451&0.506\\\hline
Task $\times$ RSI &0.390&0.195&2,36&3.169&0.054\\\hline
\boldmath$N = 64$ &&&&&\\\hline
Task &0.388&0.194&2,36&9.529&\boldmath$< 0.001$\\\hline
RSI &0.028&0.028&1,36&1.384&0.247\\\hline
Task $\times$ RSI &0.226&0.113&2,36&5.549&\textbf{0.008}\\\hline
\boldmath$N = 128$ &&&&&\\\hline
Task &0.125&0.062&2,36&3.189&0.053\\\hline
RSI &0.041&0.041&1,36&2.103&0.156\\\hline
Task $\times$ RSI &0.321&0.160&2,36&8.205&\textbf{0.001}\\\hline
\boldmath$N = 256$ &&&&&\\\hline
Task &0.157&0.079&2,36&5.538&\textbf{0.008}\\\hline
RSI &0.003&0.003&1,36&0.243&0.625\\\hline
Task $\times$ RSI &0.249&0.124&2,36&8.752&\textbf{0.001}\\\hline
\boldmath$N = 512$ &&&&&\\\hline
Task &0.194&0.097&2,36&6.628&\textbf{0.004}\\\hline
RSI &0.035&0.035&1,36&2.398&0.130\\\hline
Task $\times$ RSI &0.131&0.065&2,36&4.465&\textbf{0.019}\\\hline
\boldmath$N = 1020$ &&&&&\\\hline
Task &0.159&0.080&2,30&5.618&\textbf{0.008}\\\hline
RSI &0.044&0.044&1,30&3.113&0.088\\\hline
Task $\times$ RSI &0.208&0.104&2,30&7.354&\textbf{0.003}\\\hline
\end{tabular}
\noindent{}\\
\noindent{}$^{*}$Boldfaced values indicate significant differences at the two-tailed alpha of 0.05.

\noindent{}\\

In contrast to the HK method, the effects of task constraints $\hat{H}$ estimated using the DFA again vary as a function of RT time series length (\textbf{Fig. \ref{fig:DFAvsHKp_13}}; \textbf{Table 6}). For $N = 32$, the $\hat{H}$ values neither varied with Task nor with RSI. For $N = 128$, neither Task nor RSI affects the $\hat{H}$ values estimated using the DFA, but the two factors show a significant interaction effect. For $N = 64, 256, 512, 1020$, the $\hat{H}$ values estimated using DFA show similar sensitivity to the task constraints as do the $\hat{H}$ values estimated using the HK method.

\begin{figure*}
\centering
\includegraphics[width=0.75\textwidth]{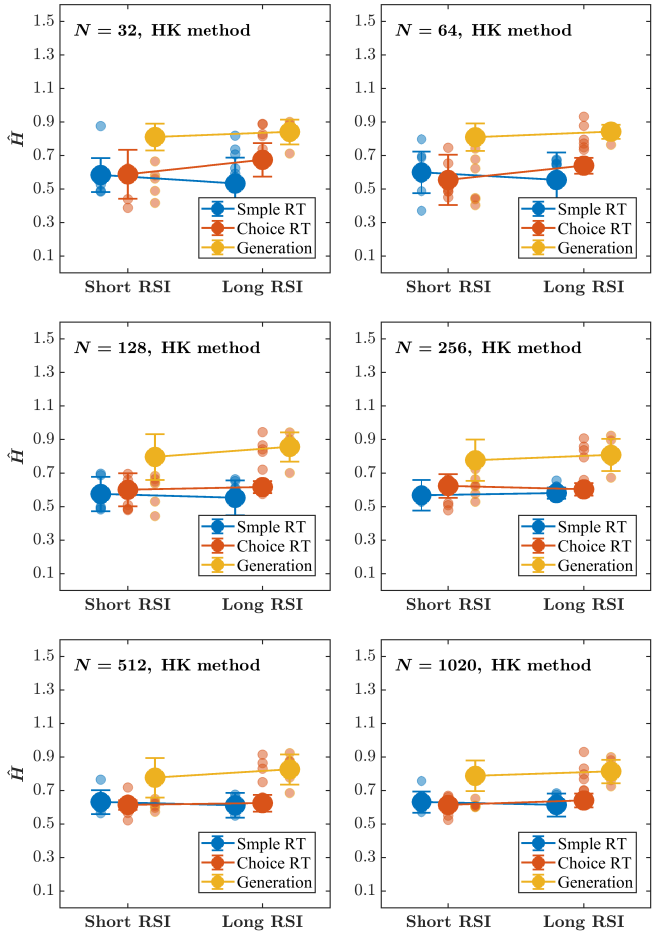}
\caption{\textbf{The effects of Task and RSI on the Hurst exponent, \boldmath$\hat{H}$, estimated using the HK method do not depend on the response-stimulus interval time series length (see Table 5 for the outcomes of the statistical tests).} Each panel plots the $Mean$ values of $\hat{H}$, estimated using the HK method for the response-stimulus interval time series of length $N = 32, 64, 128, 256, 512, 1020$. Light blue and light red circles indicate $\hat{H}$ values for individual participants in the respective conditions. Error bars indicate $95$\% CI across $6$ participants.}
\label{fig:DFAvsHKp_10}
\end{figure*}

These results further support our proposition that adopting the HK method might reduce the likelihood of the Type II error in not being able to find an actual effect of task constraints on the Hurst exponent when it exists---irrespective of whether the task is predominantly motor (e.g., walking, running) or predominantly cognitive (e.g., simple RT, complex RT).

\begin{figure*}
\centering
\includegraphics[width=0.75\textwidth]{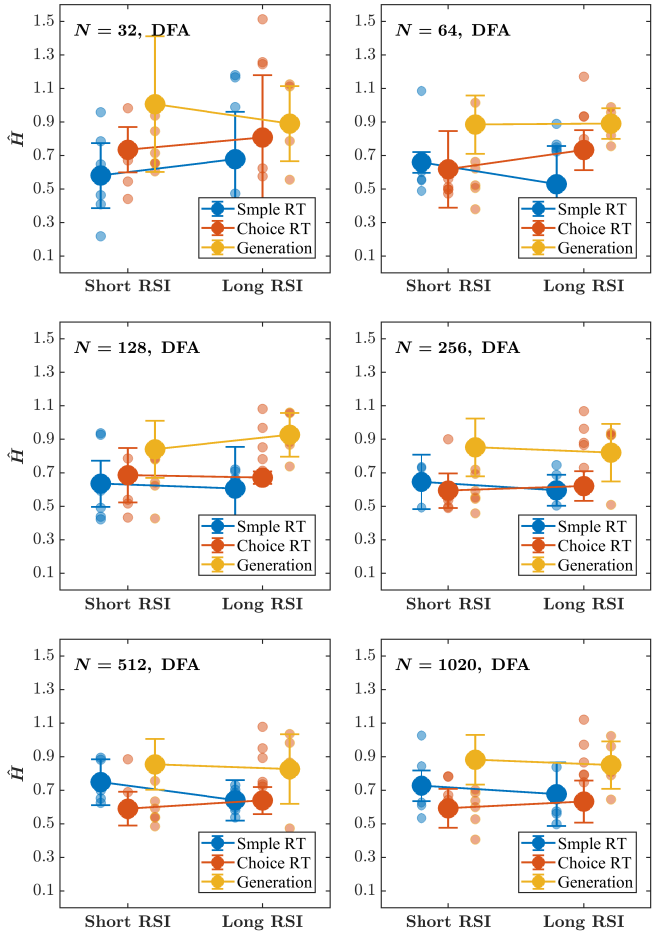}
\caption{\textbf{The effects of Task and RSI on the Hurst exponent, \boldmath$\hat{H}$, estimated using DFA wax and wane depending on the response-stimulus interval time series length (see Table 6 for the outcomes of the statistical tests).} Each panel plots the $Mean$ values of $\hat{H}$, estimated using DFA for the response-stimulus interval time series of length $N = 32, 64, 128, 256, 512, 1020$. Light blue and light red circles indicate $\hat{H}$ values for individual participants in the respective conditions. Error bars indicate $95$\% CI across $6$ participants.}
\label{fig:DFAvsHKp_11}
\end{figure*}

\subsection{Context 4: RT time series in a listening task}

The above examples of stride interval time series, tapping interval time series, and RT time series illustrate that the HK method reduces the likelihood of Type II error when comparing the Hurst exponent across task constraints from multiple measurement modalities and settings, ranging from gross and fine motor tasks to classic cognitive/psychological experiments. However, other things equal, Types I and II errors are inversely related \cite{cohen1982or,degroot2012probability,harrison2013minimizing,walley2021optimising}, raising the possibility that the HK method might increase the likelihood of Type I error, i.e., increasing the likelihood of finding the effect of an independent factor when it does not exist. To investigate whether this is the case, we analyzed an RT time series dataset in which the $H$ values estimated using DFA did not differ as a function of task constraints.

\subsubsection{Methods}

RT time series were reanalyzed from a published study \cite{bloomfield2021perceiving}. Data was collected on twenty adults (nine men and eleven women, $M \pm SD$ age = $20.10 \pm 1.29$ years) after obtaining informed consent. Participants were randomly assigned to hear the voice of either an adult woman or Acapela’s U.S. English text-to-speech female voice ``Sharon" (Acapela Inc., Mons, Belgium), using the iPad app ``Voice Dream." They both produced speech recordings of 2,027 words from \textit{The Atlantic} article ``Torching the Modern-Day Library of Alexandria." Also, they both produced words interspersed with pauses to allow parsing.
 
E-Prime software (Psychology Software Tools Inc., Pittsburgh, PA) presented audio recordings of each word in its original sequence through headphones. Participants sat at an E-Prime-ready computer and were instructed: ``Listen to the audio stimuli and press the spacebar after you feel as though you have understood the word you just heard. Try to pay attention to the passage because comprehension and word-memory questions will be asked at the end of the experiment. However, if you miss a word, do not worry and continue to move on because you cannot go back.” When the spacebar was released, a word recording played. The following word may be heard by pressing the spacebar once more, either during or after the recording playback, allowing participants to skip the entire word in favor of the subsequent one. E-Prime measured the RT in ms from the beginning of each word until the succeeding button press. The experiment yielded RT time series of lengths $N = 2027$ and $2025$ in human speech and text-to-speech conditions, respectively. RT time series of lengths $N = 32, 64, 128, 256, 512, 1024$ were submitted to the HK method and DFA. Intertap interval time series of all six lengths were shuffled to preserve the probability distribution but destroyed any temporal correlations and submitted to the HK method and DFA. As opposed to the original time series expected to yield $\hat{H} > 0.5$. these shuffled time series were expected to yield an $\hat{H}$ value of $0.5$, indicating an absence of long-range correlations.

We utilized independent samples $t$-tests to examine the effects of Speech (Human speaker vs. Text-to-speech synthesizer) on $\hat{H}$ values estimated using both methods. All tests were performed in \texttt{R} \cite{team2013r} using the function \texttt{t.test()}. Statistical significance was set at the Type I error rate of $5\%$.

\subsubsection{Results}

\begin{figure*}
\centering
\includegraphics[width=0.75\textwidth]{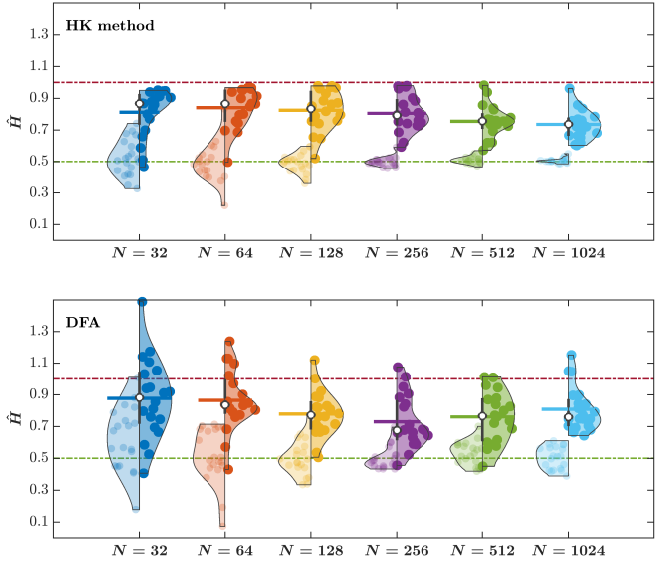}
\caption{\textbf{The Hurst exponent, \boldmath$\hat{H}$, for reaction time series estimated using the HK method do not depend on the time series length \boldmath$N$, but \boldmath$\hat{H}$ estimated using DFA show a strong dependence on \boldmath$N$, resulting in larger \boldmath$\hat{H}$ for smaller and larger \boldmath$N$.} The right and the left violin plots represent the distribution of $\hat{H}$ for the original and shuffled stride interval time series, respectively, estimated using the HK method (top) and DFA (bottom). Vertical lines represent the interquartile range of the original $\hat{H}$ values, white circles represent the median value of $\hat{H}$, and horizontal lines represent the $Mean$ value of $\hat{H}$ for the original stride interval time series. Horizontal dash-dotted green and red lines indicate $\hat{H} = 0.5$ and $\hat{H} = 1$, respectively.}
\label{fig:DFAvsHKp_12}
\end{figure*}

The central tendencies—$Mean$ and $Median$---of $\hat{H}$ for RT time series in the listening task estimated using the HK method, as well as the width of the distribution of $\hat{H}$, show a marginal reduction with the time series length $N$ (\textbf{Fig. \ref{fig:DFAvsHKp_12}, top}). In contrast, the $Mean$ and $Median$ $\hat{H}$ for RT time series estimated using DFA show a strong dependence on $N$, resulting in larger $\hat{H}$ for smaller and larger $N$ (\textbf{Fig. \ref{fig:DFAvsHKp_12}, bottom}). Furthermore, while the $\hat{H}$ values estimated using the HK method lie with the tight bounds of $[0,1]$, the $\hat{H}$ values estimated using the DFA frequently exceed the upper bound of $1$, especially for short time series. Another notable distinction is a visibly narrower range of $\hat{H}$ for the shuffled RT time series estimated using the HK method compared to DFA. As we observed in the above-discussed examples, the HK method estimates $\hat{H}$ that show smaller dispersion about the central tendency and lesser dependence on the length of the RT time series.

To investigate whether the high task sensitivity of the Hurst exponent estimated using the HK method---as shown in the above-described examples---can result in a Type I error, we next analyzed examine the effects of Speech on $\hat{H}$ values estimated using both the HK method and DFA. We submitted the $\hat{H}$ values estimated using both methods to independent samples $t$-tests. We performed these tests separately for each time series length $N = 32, 64, 128, 256, 512, 1024$.

\begin{figure*}
\centering
\includegraphics[width=0.75\textwidth]{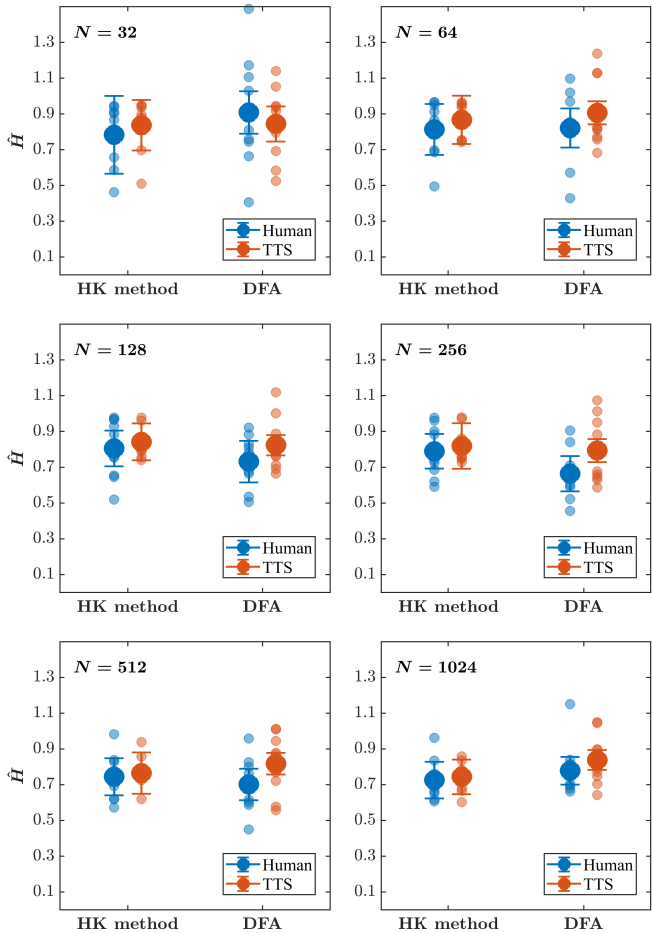}
\caption{\textbf{The effect of the speaker---human vs. text-to-speech (TTS) synthesizer---on the Hurst exponent, \boldmath$\hat{H}$, estimated using the HK method and DFA does not depend on the reaction time series length.} Each panel plots the $Mean$ values of $\hat{H}$, estimated using the HK method and DFA for reaction time series of length $N = 32, 64, 128, 256, 512, 983$. Light blue and light red circles indicate $\hat{H}$ values for individual participants in the respective conditions. Error bars indicate $95$\% CI across $10$ participants.}
\label{fig:DFAvsHKp_13}
\end{figure*}

For each time series length, the $H$ values measuring the strength of persistence in RT time series in the listening task do not differ between participants listening to the Human speaker and the Text-to-speech synthesizer, irrespective of whether these were estimated using the HK method ($t_{\textrm{9}} = -0.728, -0.941, -0.649, -0.518, -0.434, -0.425$, $p = 0.485, 0.371, 0.533, 0.617, 0.675, 0.681$ for $N = 32, 64, 128, 256, 512, 1024$, respectively) or the DFA ($t_{\textrm{9}} = 0.580, -1.124, -1.765, -1.640, -1.486, -0.967$, $p = 0.576, 0.290, 0.111, 0.135, 0.171, 0.359$ for $N = 32, 64, 128, 256, 512, 1024$, respectively; \textbf{Fig. \ref{fig:DFAvsHKp_13}}). These results suggest that the HK method balances Type I and Type II errors. Furthermore, the method reduces the likelihood of the Type II error by not missing an effect of an independent factor when it exists---as illustrated by our results on stride interval time series and RT series in simple and choice RT tasks, without increasing the likelihood of the Type I error by finding an effect of an independent factor when it does not exist---as illustrated by this example.

\section{Discussion}

We compared the performance of two methods of fractal analysis---the current gold standard, DFA, and a Bayesian approach that is not well-known in behavioral sciences: the Hurst-Kolmogorov (HK) method---in estimating the Hurst exponent of synthetic and multiple empirical time series. Simulations demonstrate that the HK method consistently outperforms the DFA in three critical ways: the HK method (i) accurately assesses long-range correlations when the measurement time series is short, (ii) shows minimal dispersion about the central tendency, and (iii) yields a point estimate that does not depend on the length of the measurement time series or its underlying Hurst exponent. Comparing the two methods using empirical time series from multiple settings further supports those findings.

The practical limitations of DFA (e.g., $N \geq 500$) is a significant drawback across the board in basic, applied, and clinical areas of science \cite{marmelat2019fractal}. From a fundamental science perspective, experiments are often constructed to obtain long sequences of measurements (e.g., RTs, stride intervals, heartbeats). Those experimental designs are slow to collect, create physical and cognitive burdens for participants, and potentially confound with fatigue. Additionally, assessing the immediate influence of experimentally induced perturbations is often desirable. However, a requirement for long time series makes it difficult to determine whether observed dynamics result from immediate reaction or longer-term learning. These concerns amplify in applied and clinical domains concerned with real-time monitoring and quick clinical assessments. We show that the HK method might help bypass these limitations because the method estimates the Hurst exponent with reasonable accuracy for time series as short as $64$ samples. Specifically, we found that the Hurst exponent yielded by the HK method closely matches the \textit{a priori} known Hurst exponent of synthetic time series as short as $64$ samples. In contrast, DFA consistently overestimates the Hurst exponent for short time series and for time series with large actual Hurst exponent. Furthermore, while the difference in performance tends to shrink with increasing time series length, the HK method consistently outperformed DFA, producing a notably smaller error in estimating the Hurst exponent even for time series as long as $1052$ samples.

Numerous authors have noted that DFA produces a large dispersion around the $Mean$ estimate of the Hurst exponent and that this dispersion increases with the actual Hurst exponent and decreases with the time series length \cite{almurad2016evenly,delignieres2006fractal,marmelat2019fractal,ravi2020assessing,roume2019biases,yuan2018unbiased}, factors that may severely limit the reproducibility of research findings. Alterations to the DFA algorithm, for instance, the evenly spacing algorithm used in our simulations and subsequent analyses, reduce the dispersion around the $Mean$ by as much as $36\%$ \cite{almurad2016evenly}. Our simulations show that the HK method can estimate $H$ with almost no dispersion around the $Mean$ estimate for time series as short as $64$ samples. Even for time series of just $32$ samples, the dispersion is negligible, as opposed to considerable dispersion in the Hurst exponent estimated by DFA. Hence, the HK method confers substantial benefits over the traditional DFA by increasing estimates' consistency---a critical feature when using the Hurst exponent as a biomarker in clinical applications wherein the objective is to differentiate between groups (e.g., healthy vs. pathological individuals).

Finally, the present results provide irrefutable evidence that while DFA is precariously sensitive to the time series length---as has been known for long \cite{delignieres2006fractal,stroe2009estimating}, the HK method yields consistent values of the Hurst exponent irrespective of the time series length. For instance, in one study \cite{marmelat2019fractal}, the Hurst exponent derived from the first $150$ strides of the 15-min walking experiment did not match the Hurst exponent from the entire 15-min trial. We also found comparable trends with the Hurst exponent estimated by DFA for empirical data on stride-to-stride variations for walking and running both on the treadmill and the overground surface, but the Hurst exponent estimated by the HK method remained consistent across different lengths taken from the empirical time series.

Empirical data poses several issues that might influence the accuracy and dispersion in the estimation of the Hurst exponent, such as trends \cite{bryce2012revisiting,hu2001effect,horvatic2011detrended}, nonstationarity \cite{bryce2012revisiting,chen2002effect}, nonlinearity \cite{chen2005effect}, and the Hurst exponent being larger than one \cite{bashan2008comparison,carpena2021validity,gao2012culturomics,telesca2010long}. Therefore, multiple efforts have been made to tailor the DFA algorithm to make it more suitable for empirical data showing one or more of these issues \cite{gao2011facilitating,shang2009chaotic,nagarajan2005minimizing,nagarajan2005minimizingDFA,nagarajan2005minimizingDFAper,qian2011modified,xu2009minimizing}. Future studies could investigate how the HK method is sensitive to the presence of either one or a combination of strong trends, nonstationarity, nonlinearity, and larger-than-one $H$. Our research team is currently involved in all those aspects.

\section{Conclusion}

The purpose of the work presented above was to compare the HK method and DFA in several contexts relevant to behavioral scientists interested in time series analysis. Without variation, simulation results showed that the HK method bypasses many of the known limitations of DFA; it (i) accurately assesses long-range correlations when the measurement time series is short, (ii) shows minimal dispersion about the central tendency, and (iii) yields a point estimate that does not depend on the length of the measurement time series or its underlying Hurst exponent. In contrast, our results also show that the DFA results applied to brief measurement time series $(N \leq 500$ should be interpreted with caution. As a general conclusion, the HK method outperforms DFA in many ways, encouraging its systematic application to assess the strength of long-range correlations in empirical time series in behavioral sciences.

\backmatter

\bmhead{Acknowledgments}

This work was supported by the NSF award 212491, the University of Nebraska Collaboration Initiative, the Center for Research in Human Movement Variability at the University of Nebraska at Omaha, the NIH awards P20GM109090 and R01NS114282, the NASA EPSCoR mechanism, and the IARPA WatchID award.

\bmhead{Author contributions}

Conceptualization: A.D.L, M.M., A.Y.W., A.C., C.M.; Methodology: A.D.L., M.M., A.Y.W., A.C., C.M.; Formal analysis: A.D.L. and M.M.; Data curation: A.D.L., M.M., A.Y.W., A.C., C.M.; Writing -- Original Draft: A.D.L. and M.M.; Writing -- Review \& Editing: A.D.L., M.M., A.Y.W., A.C., C.M.; Visualization: M.M.; Supervision: A.D.L.; Project administration: A.D.L.; Funding acquisition: A.D.L.

\bmhead{Declarations}

The authors declare no competing financial interests.

\bibliography{sn-bibliography}

\end{document}